\documentclass[pdftex,twocolumn,epjc3]{svjour3}
\usepackage{mathptmx}
\usepackage{amsmath,amssymb,multirow,graphicx}
\usepackage[numbers,sort&compress]{natbib}
\usepackage[colorlinks,citecolor=blue,urlcolor=blue,linkcolor=blue]{hyperref}

\journalname{Eur. Phys. J. C}

\title{Constraints on the braneworld from compact stars}

\author{R.~Gonz\'{a}lez Felipe\thanksref{e1,addr1,addr2}
\and
D. Manreza Paret\thanksref{e2,addr3}
\and
A. P\'{e}rez Mart\'{\i}nez\thanksref{e3,addr4,addr5}
}

\thankstext{e1}{e-mail: ricardo.felipe@tecnico.ulisboa.pt}
\thankstext{e2}{e-mail: dmanreza@fisica.uh.cu}
\thankstext{e3}{e-mail: aurora@icimaf.cu}

\institute{ISEL - Instituto Superior de Engenharia de Lisboa, 
	Instituto Polit\'ecnico de Lisboa, 
	Rua Conselheiro Em\'{\i}dio Navarro 1959-007 Lisboa, 
	Portugal\label{addr1}
	\and
	Departamento de F\'{\i}sica and Centro de F\'{\i}sica Te\'{o}rica de Part\'{\i}culas - CFTP, Instituto Superior T\'{e}cnico,
	Universidade de Lisboa, Avenida Rovisco Pais, 1049-001 Lisboa,
	Portugal\label{addr2}
	\and
	Departamento de F\'{\i}sica General, Facultad de F\'{\i}sica, 
	Universidad de la Habana, La Habana, 10400, Cuba\label{addr3}
	\and
	Instituto de Cibern\'{e}tica, Matem\'{a}tica y F\'{\i}sica (ICIMAF), 
	La Habana, 10400, Cuba\label{addr4}
	\and
	Instituto de Ciencias Nucleares, Universidad Nacional Aut\'{o}noma de M\'{e}xico,
	Apartado Postal 70-543, Distrito Federal 04510, Mexico\label{addr5}
}

\date{Received: date / Accepted: date} % The correct dates will be entered by the editor

\begin{document}
	
	\maketitle
	
	\begin{abstract}
	According to the braneworld idea, ordinary matter is confined on a 3-dimensional space (brane) that is embedded in a higher-dimensional space-time where gravity propagates.  In this work, after reviewing the limits coming from general relativity, finiteness of pressure and causality on the brane, we derive observational constraints on the braneworld parameters from the existence of stable compact stars. The analysis is carried out by solving numerically the brane-modified Tolman-Oppenheimer-Volkoff equations, using different representative equations of state to describe matter in the star interior. The cases of normal dense matter, pure quark matter and hybrid matter are considered.
	\end{abstract}

\section{Introduction}
\label{sec:intro}

Braneworld ideas and, in particular, the Randall-Sundrum (RS) models~\cite{Randall:1999ee,Randall:1999vf} have been extensively investigated during the last decade, mainly motivated by the development of string theory. A remarkable feature of the braneworld is that it modifies the Einstein equations locally and non-locally, leading to an effective energy-momentum tensor. These modifications have important consequences for cosmology~\cite{Maartens:2010ar}, since significant deviations from  Einstein gravity could have occurred at very high energies in the early universe. In particular, in brane cosmology, the expansion rate of the universe $H$ scales with the energy density $\rho$ as $H\propto \rho$, whereas this dependence is $H\propto\rho^{1/2}$ in standard cosmology. This high-energy behaviour, which is generic and not specific to RS braneworld scenarios, may affect early universe phenomena, such as inflation~\cite{Maartens:1999hf} and the generation of the cosmological baryon asymmetry~\cite{Bento:2004pz,Bento:2005xk,Okada:2005kv,Bento:2005je}.

In the Randall-Sundrum type-II brane model, the bulk geometry is curved and the brane is endowed with a tension that is fine-tuned against the bulk cosmological constant to ensure a flat Minkowski space-time in the brane. The brane tension $\lambda$ relates the Planck masses, $M_P$ and $M_5$, in four and five dimensions, respectively, via the equation $\lambda=3M_5^6/(4 \pi M_P^2)$, where $M_P=1.22 \times 10^{19}$~GeV. Successful big bang nucleosynthesis requires that the change in the expansion rate due to the new terms in the Friedmann equation be sufficiently small at scales $\sim O$(MeV). A more stringent bound can be obtained by requiring the theory to reduce to Newtonian gravity on scales larger than 1 mm~\cite{Maartens:1999hf}.

Modifications to Einstein gravity are also relevant in the vicinity of massive compact objects as black holes and neutron (quark, hybrid) stars. Compact stars are therefore a special laboratory to look for possible modifications of general relativity (GR) and to test extra dimensions~\cite{Psaltis:2008bb}. From the existence and stability of such astrophysical objects, one expects additional constraints on the parameters of alternative theories of gravity~\cite{Germani:2001du,Deruelle:2001fb,Pani:2011xm,Ovalle:2013vna,Castro:2014xza,Garcia-Aspeitia:2014pna,Linares:2015fsa,Lugones:2015}. In particular, it has been shown that neutron stars (NS) put a lower bound on the brane tension $\lambda$, which is stronger than the bound coming from big bang nucleosynthesis, although weaker than the experimental Newton law limit~\cite{Germani:2001du}. Furthermore, the well-known compactness limit $GM/R\leq 4/9$~\cite{Schwarzschild:1916ae}, obtained in GR by requiring the finiteness of pressure at the centre of a uniform star, is reduced by high-energy 5D gravity effects.

The macroscopic properties of compact stars also crucially depend on the constituent matter in the star interior. More precisely, star masses and radii are determined by the equation of state (EoS) of matter, i.e. by the relation $p(\rho)$ between the pressure and the energy density inside the star. In the absence of a unique framework to describe the physics of compact stars, several approaches can be adopted for the determination of the EoS. For instance, it can be reconstructed from mass-radius measurements~\cite{Lattimer:2006xb,Ozel:2010fw,Steiner:2010fz}. Alternatively, one can resort to theoretical calculations based on chiral effective field theories and obtain stringent constraints for the NS radii~\cite{Hebeler:2010jx}, when combined with mass measurements.

In our analysis, the non-local ``dark" components~\cite{Maartens:2010ar,Ovalle:2013vna} arising from the bulk Weyl tensor are modelled via the simple linear proportionality relation $\mathcal{P}=w\,\mathcal{U}$ between the dark energy $\mathcal{U}$ and dark pressure $\mathcal{P}$. Such a functional form follows, for instance, from the requirement that the vacuum on the brane admits a one-parameter group of conformal motions and  the field equations are invariant with respect to the Lie group of homologous transformations~\cite{Harko:2004ui}. The above relation has also been used in~\cite{Deruelle:2001fb} to study the junction conditions between the interior and exterior of static and spherically symmetric stars on the brane. Recently, it has been employed in the study of the mass-radius relation of some hadronic stars, hybrid stars (HS) and quark stars (QS) in the braneworld~\cite{Castro:2014xza}. In a different context, this type of state-like relation is commonly assumed in cosmology to describe the matter content of the universe at different epochs ($w=1/3$ for radiation domination, $w=0$ for a matter dominated universe,  and $w=-1$ for a cosmological constant).

It is pertinent to stress the limitations of our approach. We admittedly consider a linear relation between the dark energy $\mathcal{U}$ and dark pressure $\mathcal{P}$ that appear in the effective 4D equations. These quantities come from the 5D Weyl tensor, which cannot be determined solely from the 4D equations. The particular choice $\mathcal{P} = w \mathcal{U}$ is therefore a restriction. The alternative ways to proceed would be either to consider an infinite number of functional dependences $\mathcal{P}(\mathcal{U})$ or, more appropriately, to remove the ambiguity by solving the full 5D Einstein equations. The latter has proven a very demanding task so far.

The purpose of the present work is two-fold. First we study the limits coming from general relativity, finiteness of pressure and causality, applied to compact objects in the braneworld. In particular, assuming a star interior with a vanishing dark pressure and a non-vanishing dark density, and considering a pure causal EoS, we shall derive a brane-modified causality limit that depends on the brane tension, and which is more restrictive as $\lambda$ decreases. Secondly, we study compact stars in the RS type-II braneworld in order to establish limits on the parameters of the model.

Our analysis is based on astrophysical observations and we use representative EoS to describe matter in the star interior. In particular, for dense nuclear matter, we shall consider the analytical representation given in~\cite{Potekhin:2013qqa} for the unified Brussels-Montreal EoS models, which are based on the nuclear energy-density functional theory with generalized Skyrme effective forces. For quark matter, we shall employ the simple phenomenological parametrisation given in~\cite{Alford:2004pf}, which includes QCD and strange quark mass corrections. We shall also consider a hybrid EoS to study hybrid stars, i.e., stars with a hadronic outer region surrounding a quark (or mixed hadron-quark) inner core. 

The study of the macroscopic stellar properties is carried out through the solution of the Tolman--Oppenheimer-Volkoff (TOV) equations, properly modified to include local and non-local bulk effects.  The local and non-local bulk corrections to the energy-momentum tensor turn out to play a crucial role in the stability of compact stars on the brane and in establishing agreement with the observational constraints.  The present work extends previous studies (e.g.~\cite{Castro:2014xza}) not only by including in the analysis the compactness and causality limits in the braneworld, but also by considering the full parameter space for the brane tension $\lambda$ and the $w$ coefficient of the assumed dark EoS.

The paper is organised as follows. In Sect.~\ref{sec:tov} we present and briefly discuss the TOV equations on the brane. The brane-modified compactness limits are summarised in Sect.~\ref{sec:limits} and a new causality limit is derived requiring the subluminality of the EoS for matter inside the compact star. In Sect.~\ref{sec:solutions}, we present the brane TOV solutions for several representative EoS. The predicted mass-radius relations are then compared with the observational constraints. Finally, our concluding remarks are given in Sect.~\ref{sec:conclusions}.

\section{TOV equations on the brane}
\label{sec:tov}

The field equations induced on the brane have the form~\cite{Maartens:2010ar}
\begin{equation}\label{Gmunu}
G_{\mu \nu}=R_{\mu \nu}-\frac{1}{2}R g_{\mu \nu}=\kappa^2 T_{\mu \nu} + \frac{6\kappa^2}{\lambda}\,\mathcal{S}_{\mu \nu}-\mathcal{E}_{\mu \nu}\,,
\end{equation}
where $G_{\mu \nu}$ is the usual Einstein tensor, $ \kappa^2=8\pi G $, and $T_{\mu \nu}$ is the standard energy-momentum tensor.\footnote{Hereafter, we use a system of natural units with $c=1$.} For a spherically symmetric star, the metric in static coordinates is given by
\begin{equation}
ds^2=-e^{2\Phi(r)} dt^2+e^{2\Lambda(r)} dr^2+ r^2\, (d\theta^2+\sin^2\theta\, d\phi^2).
\end{equation}
The tensors $\mathcal{S}_{\mu \nu}$ and $\mathcal{E}_{\mu \nu}$ encode the local and non-local bulk corrections, respectively. For a perfect fluid, the expressions for $T_{\mu \nu}$ and $\mathcal{S}_{\mu \nu}$ are given by~\cite{Maartens:2010ar}
\begin{equation}\label{Tmunu}
  T_{\mu \nu} = \rho u_{\mu}u_{\nu}+p\, (g_{\mu \nu}+u_{\mu}u_{\nu})
\end{equation}
and
\begin{equation}\label{Smunu}
  \mathcal{S}_{\mu \nu} = \frac{1}{12}\,\rho^{2}u_{\mu}u_{\nu}+\frac{1}{12}\,
\rho(\rho+2p)(g_{\mu \nu}+u_{\mu}u_{\nu}),
\end{equation}
where $u^{\mu}$ is the four-velocity of the fluid. The tensor $\mathcal{E}_{\mu \nu}$ reduces to the form
\begin{equation}\label{Emunu}
\begin{split}
    \mathcal{E}_{\mu \nu}=&-\frac{6}{\kappa^{2}\lambda}\,\left[\mathcal{U}u_{\mu}u_{\nu}+
    \mathcal{P} r_{\mu}r_{\nu}\right.\\
    &\left.+\frac{1}{3}(\mathcal{U}-\mathcal{P})\,(g_{\mu \nu}+u_{\mu}u_{\nu}) \right],
    \end{split}
\end{equation}
for the case of a static spherical symmetry. Here, $r_{\mu}$ is a unit radial vector,  $\mathcal{U}$ is the non-local energy density (dark radiation) and $\mathcal{P}$ is the non-local pressure (dark pressure) on the brane. From Eqs.~\eqref{Gmunu} and~\eqref{Emunu}, we see that standard 4D general relativity is recovered in the limit $\lambda\rightarrow \infty$.

Solving Einstein's equations for a perfect fluid matter, the following modified TOV equations are obtained on the brane:
\begin{eqnarray}
\frac{dm}{dr}&=&4\pi r^2\rho_{\mathrm{eff}},\label{toveq1}\\
\frac{dp}{dr}&=&-(\rho+p)\frac{d\Phi}{dr},\label{toveq2}\\
\frac{d\Phi}{dr}&=&\frac{2Gm+\kappa^2 r^3\left[p_{\mathrm{eff}}+(4\mathcal{P})/(\kappa^{4}\lambda)\right]}{2r(r-2Gm)},\label{toveq3}\\
\frac{d\mathcal{U}}{dr}&=&-\frac{1}{2}\kappa^4 (\rho+p)
\frac{d \rho}{dr}-2\frac{d\mathcal{P}}{dr}-\frac{6}{r}\mathcal{P}\nonumber\\
& &-\left(2\mathcal{P}+ 4\mathcal{U} \right)\frac{d\Phi}{dr}\,,\label{toveq4}
\end{eqnarray}
where
\begin{equation}\label{rhopeff}
  \rho_{\mathrm{eff}} = \rho_{\mathrm{loc}}+\frac{6}{\kappa^{4} \lambda}\mathcal{U}\,, \quad
  p_{\mathrm{eff}} = p_{\mathrm{loc}} +\frac{2}{\kappa^{4} \lambda}\mathcal{U} \,.
\end{equation}
In the above expressions,
\begin{equation}\label{rhoploc}
\rho_{\mathrm{loc}} = \rho+\frac{\rho^2}{2\lambda}\,, \quad
p_{\mathrm{loc}} = p+\frac{p\rho}{\lambda}+\frac{\rho^2}{2\lambda}\,,
\end{equation}
denote the effective local matter density and pressure, respectively. In order to solve the system of differential equations~\eqref{toveq1}-\eqref{toveq4}, an equation of state $p(\rho)$  for matter and a relation $\mathcal{P}(\mathcal{U})$ are required. As explained before, we shall assume the simplest state-like relation $\mathcal{P}=w\,\mathcal{U}$. With this choice, Eqs.~\eqref{toveq1}-\eqref{toveq4} become
\begin{eqnarray}
\frac{dm}{dr}&=&4\pi r^2\rho_{\mathrm{eff}},\label{toveq1a}\\
\frac{dp}{dr}&=&-(\rho+p)\frac{d\Phi}{dr},\label{toveq2a}\\
\frac{d\Phi}{dr}&=&\frac{2Gm+\kappa^2 r^3\left[p_{\mathrm{eff}}+(4w\,\mathcal{U})/(\kappa^{4}\lambda)\right]}{2r(r-2Gm)},
\label{toveq3a}\\
\frac{d\mathcal{U}}{dr}&=&-\frac{2}{1+2w}\left[\frac{\kappa^4}{4} (\rho+p)
\frac{d \rho}{dr}+ \frac{3w}{r}\mathcal{U}\right.\nonumber\\
& &\left.+(w+2)\,\mathcal{U} \frac{d\Phi}{dr}\right]\label{toveq4a},
\end{eqnarray}
where the last equation holds for $w\neq-1/2$. The case $w=-1/2$ should be treated separately. Solving for $\mathcal{U}$ in Eq.~\eqref{toveq4} and using~\eqref{toveq3}, we obtain for $w=-1/2$,
\begin{equation} \label{Uwonehalf}
\mathcal{U}(r)=\frac{\kappa^4 (\rho+p)^2}{6 v_s^2}
\frac{2Gm+\kappa^2 r^3\, p_{\mathrm{loc}}}{2(3Gm-r)+\kappa^2 r^3\, p_{\mathrm{loc}}},
\end{equation}
where $v_s$ is the speed of sound,
\begin{equation}\label{vs}
v_s^2=\frac{dp}{d\rho}.
\end{equation}

To integrate the TOV equations, we need appropriate initial conditions at the centre of the star. As in general relativity, we assume that the enclosed mass is zero at the centre, $m(0) = 0$, and that $p(0) = p_c$, where $p_c$ is the central pressure. At the stellar surface, we require $p(R) = 0$, i.e. a vanishing pressure, which corresponds to a star mass $m(R) = M$. As for the dark component $\mathcal{U}$, we shall assume that $\mathcal{U}(0) = 0$. We remark that different initial conditions for the dark energy density $\mathcal{U}$ can be chosen. They could be given either at the centre of the star or at its surface. In the latter case, a shooting method is required for the integration of the system of equations~\cite{Lugones:2015}. The boundary condition chosen here, i.e. a vanishing dark density at the centre, is the simplest choice. Note also that if $w=-1/2$, the value of $\mathcal{U}$ at the centre is obtained directly from Eq.~\eqref{Uwonehalf}.

The system~\eqref{toveq1a}-\eqref{toveq4a} must be supplemented with the Israel-Darmois junction conditions~\cite{Darmois:1927,Israel:1966rt} at the stellar surface. On the brane, this  leads to the matching condition
\begin{equation}
	\left[p_{\mathrm{eff}}+\frac{4}{\kappa^4\lambda}\mathcal{P}\right]_\text{surf}=0,
\end{equation}
where $[f]_\text{surf}\equiv f(R^+)-f(R^-)$, and the superscripts $+$ and $-$ refer to quantities defined outside and inside the star, respectively. The requirement $p(R)=0$ at the surface then implies
\begin{equation}\label{IDjunction}
	\frac{\kappa^4}{4}\rho^2(R)+\mathcal{U}^-(R)+2\,\mathcal{P}^-(R)=\mathcal{U}^+(R)+2\,\mathcal{P}^+(R).
\end{equation}

In the absence of any Weyl stress in the interior, we have $\mathcal{P}^-=\mathcal{U}^-=0$ and Eq.~\eqref{IDjunction} implies that the exterior must be non-Schwarzschild, provided that the energy density $\rho$ does not vanish at the surface. The same conclusion holds if $w=-1/2$, since in this case $\mathcal{U}^-(R)+2\,\mathcal{P}^-(R)=0$. On the other hand, if $\rho(R)\neq0$ and the exterior is Schwarzschild ($\mathcal{P}^+=\mathcal{U}^+=0$), then necessarily  $\mathcal{U}^-(R)+2\,\mathcal{P}^-(R)\neq0$ in the star interior. Note that the condition $\mathcal{P}=\mathcal{U}=0$ is the condition for a perfect fluid without non-local or Weyl corrections. While for the exterior this corresponds to the Schwarzschild solution, this is not the case for the star interior, where local high-energy corrections are present. A consistent version of the Schwarzschild interior metric in the context of the braneworld, including local and non-local bulk terms, has been found in~\cite{Ovalle:2010zc}. Regarding the matching conditions, it has been shown that the exterior Schwarzschild solution is compatible with a stellar distribution made of regular matter~\cite{Ovalle:2014uwa}.

We also note that the TOV equations in the braneworld admit asymptotically flat exterior solutions different from the Schwarzschild solution. In what follows, our analysis is restricted to solutions in the star interior with generic dark energy and dark pressure components obeying the EoS relation $\mathcal{P}=w\,\mathcal{U}$. Possible exterior solutions are not considered in this work. 

\section{Compactness limits in the braneworld}
\label{sec:limits}

In this section, we revisit the compactness limits on a uniform star coming from general relativity and the requirement of the finiteness of pressure. We shall also derive a brane-modified causality limit for the existence of stable stars.

\begin{figure}[t]
	\centering
	\hspace{-1em}\includegraphics[width=.5\textwidth]{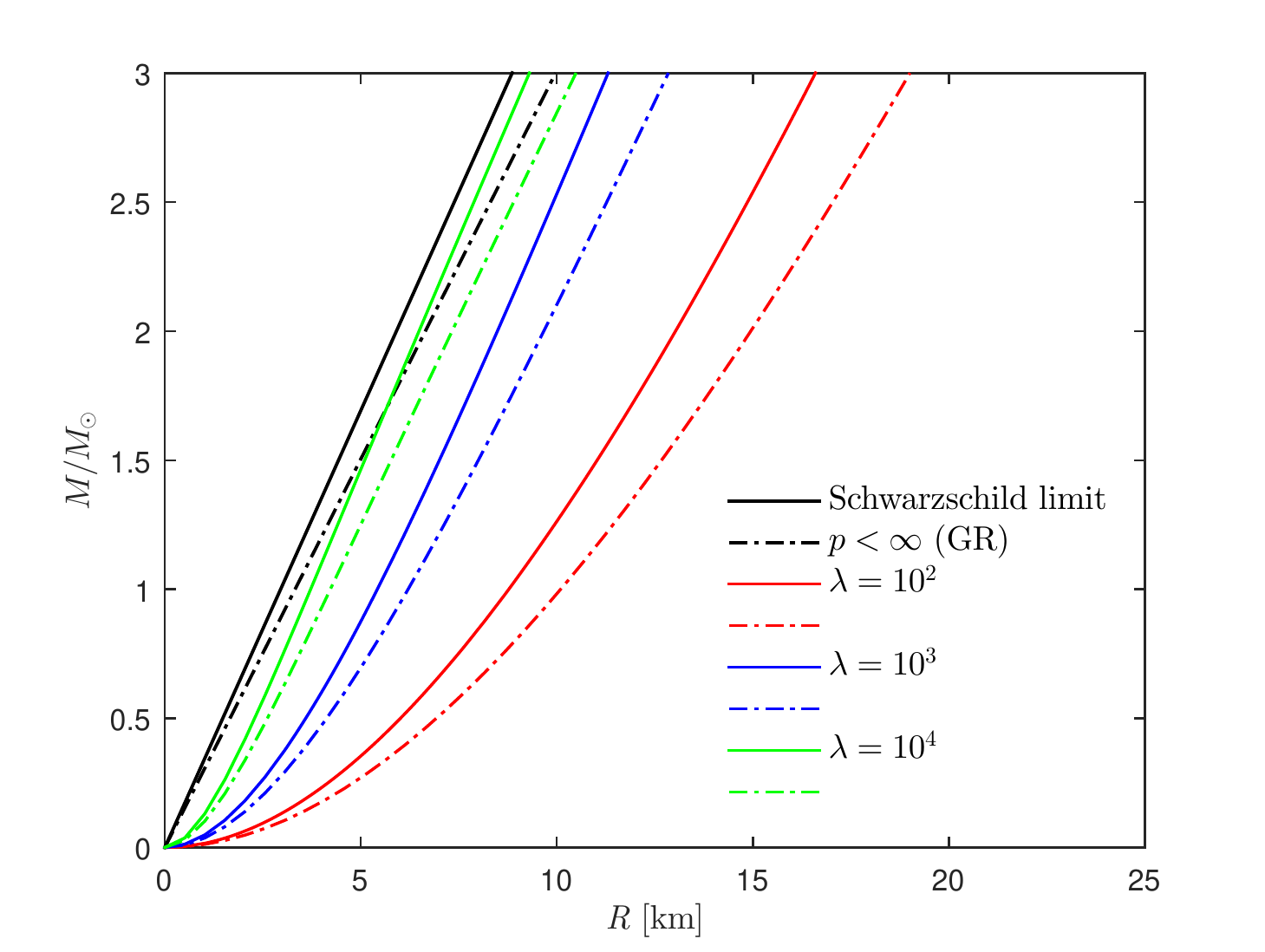}
	\caption{\label{fig1} General relativity and pressure finiteness limits on the brane, assuming a star interior with ${\cal P}={\cal U}=0$. The brane tension $\lambda$ is given in units of MeV/fm$^3$.}
\end{figure}

Assuming a uniform density and ${\cal P}={\cal U}=0$, the high-energy brane corrections are local. In this case, an astrophysical lower limit on $\lambda$, independent of the Newton-law and cosmological limits, can be established for all uniform stars~\cite{Germani:2001du}:
\begin{equation}\label{alim}
{\lambda}\geq \frac{GM\rho}{R-2GM}\,,
\end{equation}
with $\rho=3M/(4\pi R^3)$. For a positive brane tension, this implies $R>2GM$, so that the Schwarzschild radius remains a limiting radius. Below the astrophysical limit~\eqref{alim}, stable neutron stars cannot exist on the brane. This bound is much stronger than the cosmological nucleosynthesis constraint, but turns out to be weaker than the Newton-law lower bound. From Eq.~\eqref{alim}, we find the following  upper bound for the mass:
\begin{equation}\label{alim1}
M(R)=\frac{2R^2 }{3}\left(\sqrt{\pi\lambda  \left(3 G^{-1}+4 \pi  \lambda  R^2\right)}-2 \pi  \lambda R\right).
\end{equation}
As expected, in the limit $\lambda\rightarrow\infty$, we recover the Schwarzschild limit of GR:
\begin{equation}
M(R)=\frac{1}{2}\frac{R}{G}.
\end{equation}
The astrophysical bound given by Eq.~(\ref{alim1}) is illustrated in Fig.~\ref{fig1} for different values of $\lambda$ (solid lines). As can be seen from the figure, brane corrections to the compactness limit would be significant for $\lambda \lesssim 10^4$~MeV/fm$^3$, leading to a less compact star in the braneworld, when compared to the general relativity case.
	
Another important limit comes from requiring the pressure inside the compact object to be finite. Since the pressure decreases with the radius $r$, this condition is equivalent to the requirement of a positive and finite pressure at the centre of the star. This gives the constraint~\cite{Germani:2001du}
\begin{eqnarray}\label{plim}
\frac{GM}{R} \leq {4\over 9}\, \frac{1+7\rho/4\lambda +
	5\rho^2/8\lambda^2}{(1+\rho/\lambda)^2(1+\rho/2\lambda)}\,.
\end{eqnarray}
Solving for $M$ to find the maximum mass, we obtain
\begin{eqnarray} \label{plim1}
M(R)&=&\frac{4 \sqrt{\pi } R^2}{9 G} \left[-2 \sqrt{\pi } G \lambda  R\right.\nonumber\\
& & \left.+\sqrt{G \lambda  \left(5+4 \pi  G \lambda  R^2\right)}\,
	\cos \left(\frac{1}{3} \tan ^{-1}x\right)\right],
\end{eqnarray}
where
\begin{equation}
x=\frac{\left(125 G^{-1} \lambda^{-1}+264 \pi R^2+144 \pi ^2 G
		\lambda  R^4\right)^{1/2}}{2\sqrt{\pi}\, R \left(3+4\pi G \lambda  R^2\right)}.
\end{equation}
In the limit $\lambda\rightarrow\infty$, we recover the pressure finiteness limit of GR,
\begin{equation}
M(R)=\frac{4}{9}\frac{R}{G}.
\end{equation}

The curve defined by Eq.~\eqref{plim1} is presented in Fig.~\ref{fig1}  for different values of the brane tension (dot-dashed lines). As in the case of the Schwarzschild limit, high-energy braneworld corrections are significant for $\lambda \lesssim 10^4$~MeV/fm$^3$. We also note that for a given radius $R$ the maximum mass allowed by the bound~\eqref{plim1} is lower than that coming from the limit in Eq.~\eqref{alim1}.

A third relevant constraint on the maximum star mass comes from causality, i.e. from requiring the subluminality of the EoS,\footnote{For some caveats on the speed of sound and the requirement of causality see e.g.~\cite{Ellis:2007ic}.} i.e. $v_s^2\leq 1$. The bound obtained from this condition is controlled by the stiffness of the matter EoS, and several limits have been derived in the literature~\cite{Rhoades:1974fn,Glendenning:1992dr,Koranda:1996jm} (see also~\cite{Lattimer:2006xb} for a review). In particular, a stringent causal limit has been obtained in~\cite{Koranda:1996jm}, based on the ``minimum period" EoS
\begin{equation}\label{causaleos}
p=\left\{\begin{array}{ll}
0, &\quad \rho<\rho_s,\\
\rho-\rho_s,&\quad \rho\geq\rho_s,
\end{array}\right.
\end{equation}
where $\rho_s$ is the surface energy density. This EoS corresponds to maximal stiffness  at high densities and minimal stiffness at low densities, thus supporting the largest mass with the smallest radius. In general relativity, such EoS also implies that the maximum value of the mass scales with the radius as $M\propto R$. The numerical integration of the TOV equations for various initial values of the central pressure $p_c$ then leads to the GR relation~\cite{Koranda:1996jm}
\begin{equation}\label{causalR}
R\simeq 2.82\,GM.
\end{equation}
As we shall show next, this compactness limit is modified in the braneworld.

First we notice that, from Eqs.~\eqref{rhopeff} and \eqref{rhoploc}, and using the conservation equations for $\rho$ and $\mathcal{U}$, an effective speed of sound~\cite{Maartens:2010ar} is obtained as
\begin{equation}\label{vseff}
v_{s,\mathrm{eff}}^2\equiv\frac{dp_{\mathrm{eff}}}{d\rho_{\mathrm{eff}}}=\frac{v_s^2 (1+\rho/\lambda)+(\rho+p)/\lambda+v_\mathcal{U}^2}{1+\rho/\lambda+3v_\mathcal{U}^2},
\end{equation}
where
\begin{equation}
v_\mathcal{U}^2=\frac{8}{3\kappa^4\lambda}\frac{\mathcal{U}}{\rho+p}.
\end{equation}
As expected, in the limit $\lambda\rightarrow \infty$, we have $v_{s,\mathrm{eff}}=v_s$. Note also that $v_\mathcal{U}^2$ is not necessarily a positive number, since the Weyl energy density can be negative. Moreover, when the radiation term $v_\mathcal{U}$ dominates over the matter components, one has $v_{s,\mathrm{eff}}^2 \simeq 1/3$. On the other hand, if $v_\mathcal{U}^2 \ll \rho/\lambda$ then $v_{s,\mathrm{eff}}^2\simeq v_s^2+p/\rho+1$.

Let us consider the case of a Schwarzschild exterior and a star interior with ${\cal P}=0$ (i.e. $w=0$) and ${\cal U}\neq 0$. In this case, the boundary condition~\eqref{IDjunction} yields a negative dark radiation density at the star surface. We obtain ${\cal U}=-\kappa^4\rho_s^2/4 $, so that $v_\mathcal{U}^2=-2\rho_s/(3\lambda)$ at the surface. From Eq.~\eqref{vseff} we  then find
\begin{equation}\label{vseffw0}
v_{s,\mathrm{eff}}^2=\frac{v_s^2 (1+\rho_s/\lambda)+\rho_s/(3\lambda)}{1-\rho_s/\lambda}.
\end{equation}
Requiring $v_{s,\mathrm{eff}}\leq 1$, the constraint
\begin{equation}
\lambda > \frac{(4/3+v_s^2)\,\rho_s}{1-v_s^2},
\end{equation}
with $v_s^2<1$, is obtained. 

\begin{figure}[h]
	\centering
	\begin{tabular}{l}
		\hspace{-1em}\includegraphics[width=.5\textwidth]{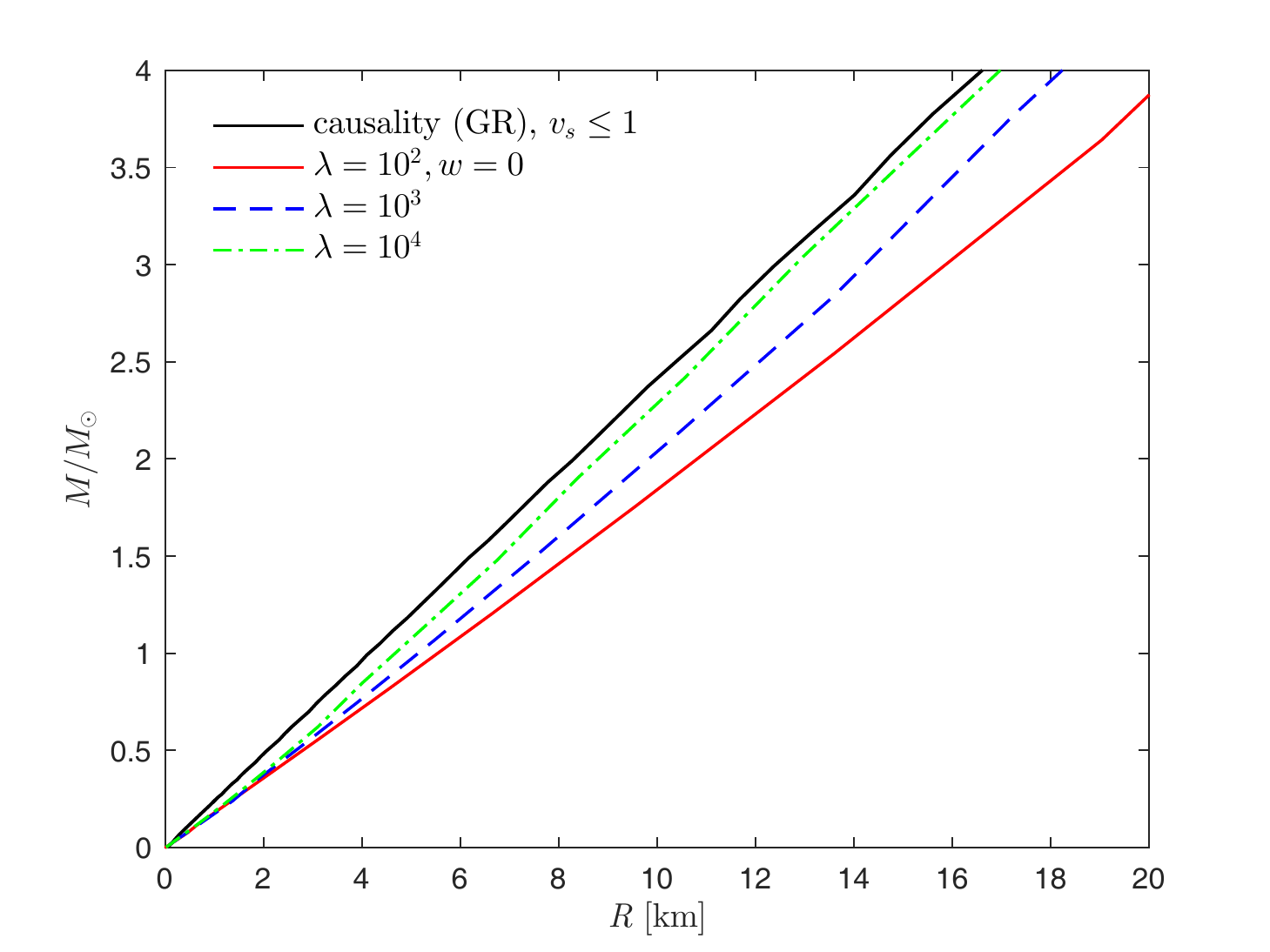}\\
		\hspace{-1em}\includegraphics[width=.5\textwidth]{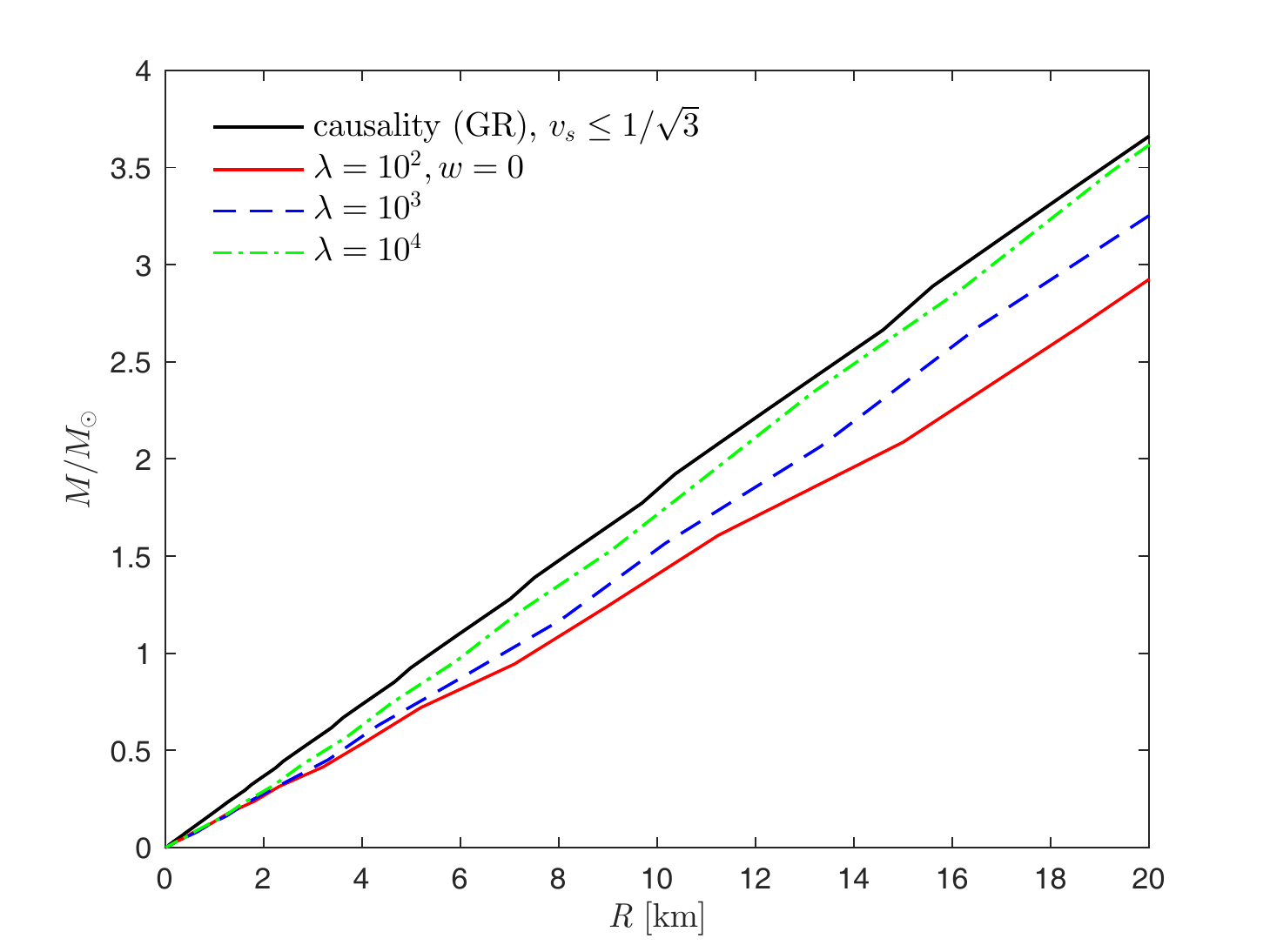}
	\end{tabular}
\caption{\label{fig2} Causality constraint on the brane, obtained from the causal EoS of Eq.~\eqref{causaleos} (top plot) and Eq.~\eqref{causaleos1} (bottom plot). A star interior with $\mathcal{P}=0$ (i.e. $w=0$) and $\mathcal{U}\neq0$ is assumed. The brane tension $\lambda$ is given in units of MeV/fm$^3$.}
\end{figure}

Next we illustrate how the corrections due to brane effects lead to modifications of the GR compactness limit given in Eq.~\eqref{causalR}. Solving numerically the system of equations \eqref{toveq1a}-\eqref{toveq4a}, from the star centre until the surface, we can determine the maximum star mass. The results are presented in Fig.~\ref{fig2} (upper plot) for different values of the brane tension $\lambda$ and taking $w=0$. Deviations from the causality limit in GR (black solid line) occur for $\lambda \lesssim 10^4$~MeV/fm$^3$, as can be seen from the curves for maximal masses depicted in the figure. In this case, for the star radii range of interest, the maximum mass (minimal radius) can be well approximated by a straight line, $M\simeq\alpha R/G$, with a slope $\alpha$ that increases as the brane tension $\lambda$ increases, reaching the GR limit~\eqref{causalR} when $\lambda\rightarrow\infty$. The values of $\alpha^{-1}$ are presented in the second column of Table~\ref{tab1} for different values of $\lambda$.

\begin{table}[h]
	\caption{\label{tab1} Causality limit in the braneworld for two causal EoS (with $v_s\leq 1$ and $v_s\leq 1/\sqrt{3}$), assuming a vanishing dark radiation pressure ($\mathcal{P}=0$) and a non-zero dark energy density $\mathcal{U}$ in the star interior. The minimum radius is approximately described by a straight line with a slope that varies with the brane tension.}
	\centering
	\begin{tabular}{ccc}
		\hline
		 & $v_s \leq 1$ &  $v_s \leq 1/\sqrt{3}$ \\
		$\lambda$ [MeV/fm$^3$]& $R/(GM)$ & $R/(GM)$\\
		\hline
		$10$ & 3.77 & 4.96\\
		$10^2$  & 3.45 & 4.64\\
		$10^3$  & 3.06 & 4.13\\
		$10^4$  & 2.83 & 3.73\\
		$\infty$ (GR)  & 2.82 & 3.68\\
		\hline
		\end{tabular}
		\end{table}
	
It has been recently conjectured that the speed of sound should satisfy the bound $v_s\leq 1/\sqrt{3}$ in any medium~\cite{Bedaque:2014sqa}. It is well known that this bound is saturated in conformal theories, including non-interacting massless gases for which $p=\rho/3$.  The limit also applies to non-relativistic and weakly coupled theories, and is respected by several strongly coupled theories. If such a bound actually holds for the speed of sound, it would modify the causality limit for compact astrophysical objects. In particular, it has been pointed out that the existence of neutron stars with $M\sim 2 M_\odot$, combined with the knowledge of the EoS of hadronic matter at low densities, is in strong tension with this bound~\cite{Bedaque:2014sqa}.

To illustrate the implications of the bound $v_s\leq 1/\sqrt{3}$ for the stability of compact stars, let us consider the minimal causal EoS
\begin{equation}\label{causaleos1}
p=\left\{\begin{array}{ll}
0, &\quad \rho<\rho_s\\
(\rho-\rho_s)/3,&\quad \rho\geq\rho_s,
\end{array}\right.
\end{equation}
which saturates the bound.\footnote{Note that the simple EoS of the well-known MIT bag model (with massless quarks and no strong coupling constant) has also this form. In this case, the parameter $\rho_s$ is associated to the bag constant, $\rho_s \equiv 4 B$.} The numerical integration of the TOV equations of GR implies the causality relation
\begin{equation}\label{causalR1}
R\simeq 3.68\,GM,
\end{equation}
which is more restrictive that the bound~\eqref{causalR}. The inclusion of brane corrections leads to modifications to this limit. For the case of a star interior with $\mathcal{P}=0$ and $\mathcal{U} \neq 0$, the results are shown in Fig.~\ref{fig2} (lower plot). Once again, significant deviations from the causality limit of GR take place for $\lambda \lesssim 10^4$~MeV/fm$^3$. As before, the minimum radius is well approximated by a straight line whose slope varies with the brane tension. The results are given in the last column of Table~\ref{tab1}. We note that, for any given value of $\lambda$, the constraint obtained from requiring $v_s\leq 1/\sqrt{3}$ is always stronger than the one previously found under the assumption $v_s\leq 1$.

\section{Brane TOV solutions and observational constraints}
\label{sec:solutions}

Due to the uncertainties in the description of the many-body interactions and the nuclear symmetry energy, as well as our lack of knowledge of the precise nature of strong interactions, the EoS of dense matter above the nuclear saturation density ($\rho_0 = 2.7\times 10^{14}\, \text{g/cm}^3 \simeq 150$~MeV/fm$^3$) is largely unknown. Depending on the matter composition in the neutron star interior, three main types of equations of state have been commonly used, namely, EoS for normal dense matter, pure quark matter or hybrid matter.

In the case of normal dense matter, neutron stars are supported against gravitational collapse by neutron degeneracy. The EoS takes into account nucleon-nucleon interactions and is characterised by a vanishing pressure at null densities. For hybrid stars, the EoS is usually softened at high densities by adding hadronic or pure quark matter at the inner core of the star, which leads to a phase transition at a given critical density. Normal and hybrid EoS do not lead to stringent bounds for the star radii, which in principle can be large ($\sim 100$~km). On the other hand, pure quark stars are conjectured to be mainly composed of strange quark matter (SQM) in the ground state, with a vanishing pressure at non-zero densities. For such stars the maximum radius is not so large ($\sim 10$~km). In all three EoS cases, the mass-radius relation predicts a maximum mass $M_{max}$. While typically $M_{max} \lesssim 2 M_\odot$ for hybrid and strange quark EoS, maximum masses up to $2.5 M_\odot$ can be reached with normal matter EoS.

From the astrophysical viewpoint, an important aspect in the study of stellar configurations is their stability. A necessary condition for the stability of a compact star is given by the so-called static criterion
\begin{equation}
\frac{dM}{d\rho_c}>0.
\end{equation}
In other words, a compact star is stable if its mass $M$ increases with growing central density $\rho_c=\rho(p_c)$. Although the stability analysis is usually quite involved, some simple criteria can be formulated based on the mass-radius configurations and their stability with respect to radial oscillations (see e.g.~\cite{Haensel:2007yy} and references therein). At each extremum (critical point) of the $M(R)$ curve, it is assumed that only one radial mode changes its stability from stable to unstable or, vice versa, from unstable to stable. Furthermore, at any critical point, a mode becomes unstable (stable) if and only if the curve bends counterclockwise (clockwise). Finally, a mode with an even (odd) number of radial nodes is said to change its stability if and only if $dR/d\rho_c>0\, (dR/d\rho_c<0)$ at the critical point. Using the above criteria, the stability of a  given stellar configuration can be easily checked.

In order to confront different EoS with current mass and radius measurements, we shall use the following observational constraints. For NS radii, we adopt the range
\begin{equation}\label{obsradii}
7.6\,{\rm km} \leq R \leq 13.9\,{\rm km},
\end{equation}
where the lower bound follows from the measurement of the NS radius using the thermal spectra from quiescent low-mass X-ray binaries inside globular clusters~\cite{Guillot:2013wu} and the upper limit is taken from the analysis of~\cite{Hebeler:2010jx}, based on the chiral effective theory. For the star masses, we consider the limits
\begin{equation}\label{obsmasses}
1.08\, M_\odot \leq M \leq 2.05\, M_\odot,
\end{equation}
where the conservative lower bound comes from the expected range for the gravitational NS birth masses~\cite{Ozel:2012ax,Kiziltan:2013oja}  and the maximum value is obtained from recent radio-timing observations of the pulsar PSR J0348+0432~\cite{Antoniadis:2013pzd}.

In the braneworld, the macroscopic properties of stable stellar configurations are controlled not only by the matter EoS but also by high-energy brane effects, characterised in our setup by the brane tension and the assumed dark EoS. Next we analyse the implications of these effects on the mass-radius relations of stable compact stars for the three classes of EoS mentioned above. In our analysis, we shall take into account the stability criteria as well as the physically plausible condition of causality $v_{s,\mathrm{eff}} \leq 1$, based on the effective speed of sound defined in Eq.~\eqref{vseff}.

\subsection{Neutron stars}

Let us first study the example of a neutron star. To describe the crust and the core of the star, we shall use an analytical representation of the Brussels-Montreal unified EoS for cold nuclear matter, referred to as models BSk19, BSk20, and BSk21~\cite{Goriely:2010bm,Pearson:2011zz,Pearson:2012hz}. We consider the following parametrisation of $p(\rho)$~\cite{Potekhin:2013qqa}
\begin{eqnarray}\label{pnseos}
\begin{split}
\zeta =&
\frac{a_1+a_2\xi+a_3\xi^3}{1+a_4\,\xi}\,
\left\{\exp\left[a_5\,(\xi-a_6)\right]+1\right\}^{-1}
\\
&+ (a_7+a_8\,\xi)\,
\left\{\exp\left[a_9\,(a_6-\xi)\right]+1\right\}^{-1}
\\
&+ (a_{10}+a_{11}\,\xi)\,
\left\{\exp\left[a_{12}\,(a_{13}-\xi)\right]+1\right\}^{-1}
\\
&+ (a_{14}+a_{15}\,\xi)\,
\left\{\exp\left[a_{16}\,(a_{17}-\xi)\right]+1\right\}^{-1}
\\
&+ \frac{a_{18}}{1+ [a_{19}\,(\xi-a_{20})]^2}
+ \frac{a_{21}}{1+ [a_{22}\,(\xi-a_{23})]^2},
\end{split}
\end{eqnarray}
where $\xi=\log_{10}(\rho/\textrm{g.cm}^{-3})$ and
$\zeta = \log_{10}(p/\text{dyn.cm}^{-2})$. The parameters $a_i$ ($i=1,\ldots,23$) for the three models are given in Table~\ref{tab2}. In what follows, we restrict our analysis to the model BSk21.

\begin{table}
	\centering
	\caption[]{Parameters $a_i$ used in Eq.~\eqref{pnseos} for the EoS BSk19, BSk20 and BSk21~\cite{Potekhin:2013qqa}.}
	\label{tab2}
	\begin{tabular}{rccc}
		\hline
		 & BSk19 & BSk20 & BSk21  \\
		\hline\rule{0pt}{2.7ex}
		$a_1$  & 3.916 &     4.078 &   4.857 \\
		$a_2$  & 7.701 &     7.587 &   6.981 \\
		$a_3$  & 0.00858 &   0.00839 & 0.00706 \\
		$a_4$  & 0.22114 &   0.21695 & 0.19351 \\
		$a_5$  & 3.269 &     3.614 &   4.085 \\
		$a_6$  & 11.964 &    11.942 &  12.065 \\
		$a_7$  & 13.349 &    13.751 &  10.521 \\
		$a_8$  & 1.3683 &    1.3373 &  1.5905 \\
		$a_9$  &  3.254 &    3.606 &   4.104 \\
		$a_{10}$ &  $-12.953$ &  $-22.996$ & $-28.726$ \\
		$a_{11}$ &  0.9237 &   1.6229 &  2.0845 \\
		$a_{12}$ &  6.20 &     4.88 &    4.89 \\
		$a_{13}$ &  14.383 &   14.274 &  14.302 \\
		$a_{14}$ &  16.693 &   23.560 &  22.881 \\
		$a_{15}$ &  $-1.0514$ &  $-1.5564$ & $-1.7690$ \\
		$a_{16}$ &  2.486 &    2.095 &   0.989 \\
		$a_{17}$ & 15.362 &    15.294 &  15.313 \\
		$a_{18}$ & 0.085 &     0.084 &   0.091 \\
		$a_{19}$ & 6.23 &      6.36 &    4.68 \\
		$a_{20}$ & 11.68 &     11.67 &   11.65 \\
		$a_{21}$ & $-0.029$ &    $-0.042$ &  $-0.086$ \\
		$a_{22}$ & 20.1 &      14.8 &    10.0 \\
		$a_{23}$ & 14.19 &     14.18 &   14.15 \\
		\hline
	\end{tabular}
\end{table}

\begin{figure}[t]
	\centering
	\begin{tabular}{l}
		\hspace{-1em}\includegraphics[width=.5\textwidth]{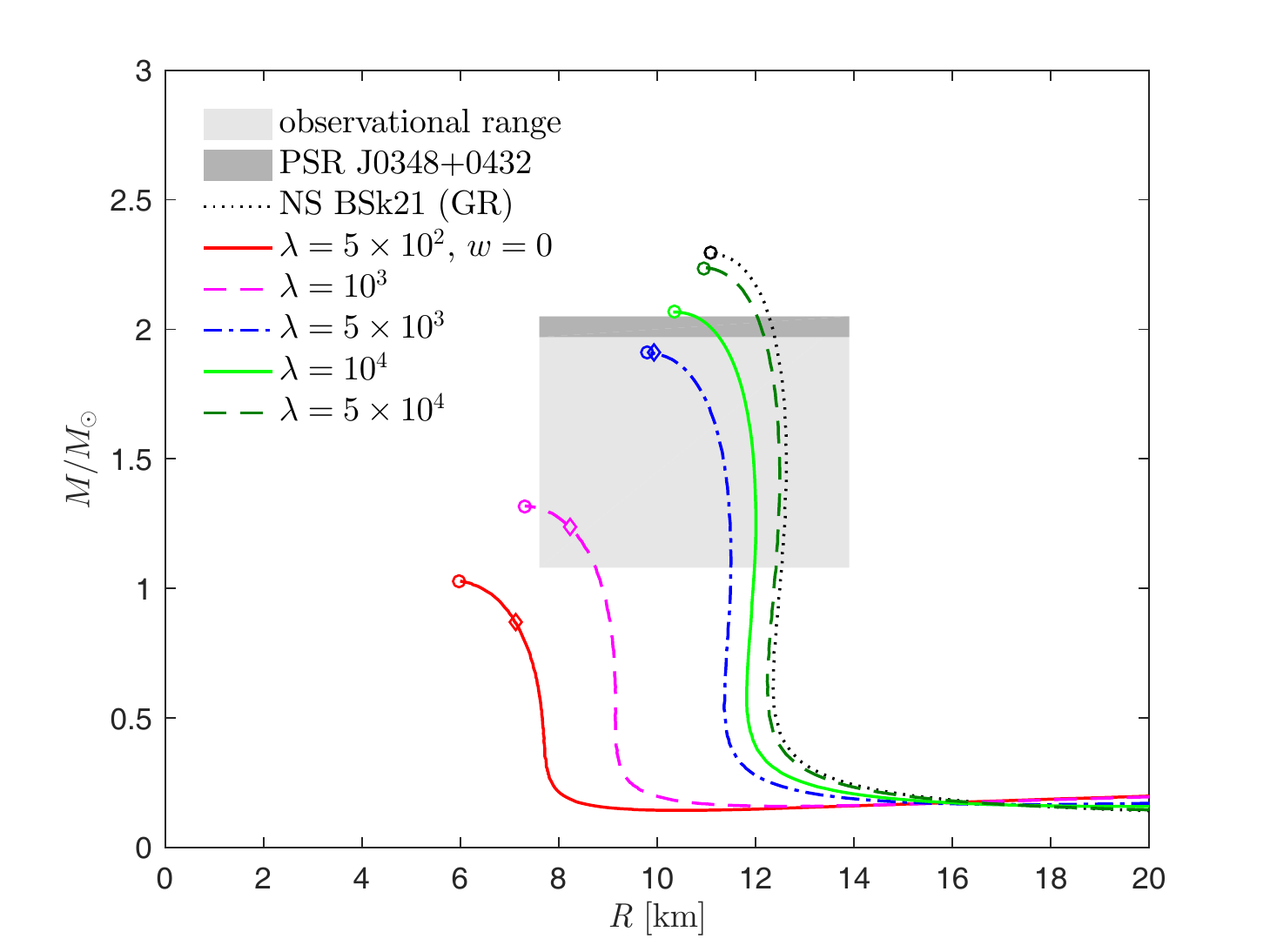}\\
		\hspace{-1em}
		\includegraphics[width=.5\textwidth]{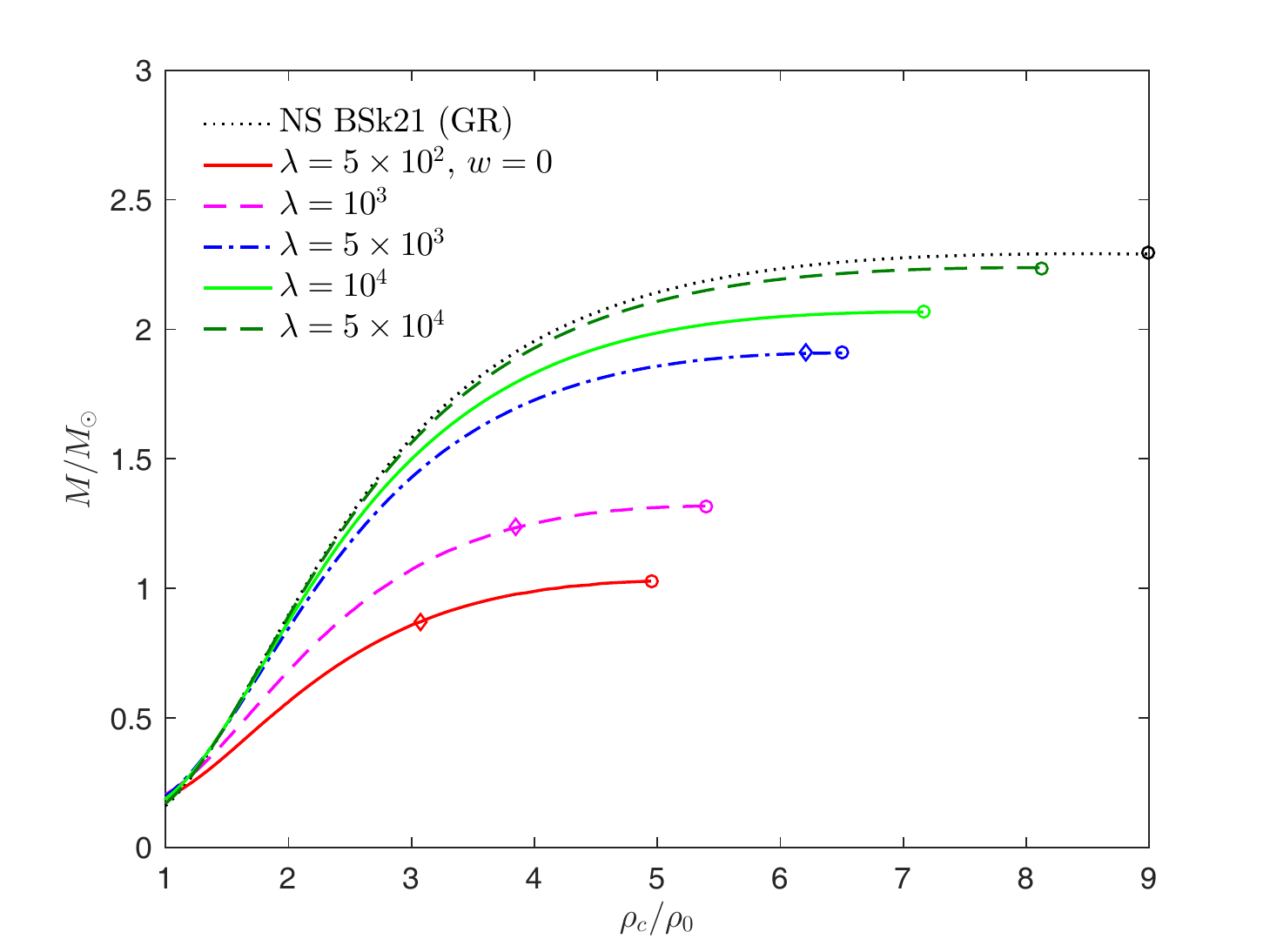}
	\end{tabular}	
	\caption{\label{fig3} Top plot: Mass-radius relation for the NS BSk21 EoS. The curves are given for different values of the brane tension $\lambda$ (in MeV/fm$^3$), assuming a vanishing dark pressure in the star interior ($w=0$) Bottom plot: The mass of the star versus the central energy density $\rho_c$ (in units of the nuclear saturation density $\rho_0$).}
\end{figure}

\begin{figure}[t]
	\centering
	\begin{tabular}{l}
		\hspace{-1em}\includegraphics[width=.5\textwidth]{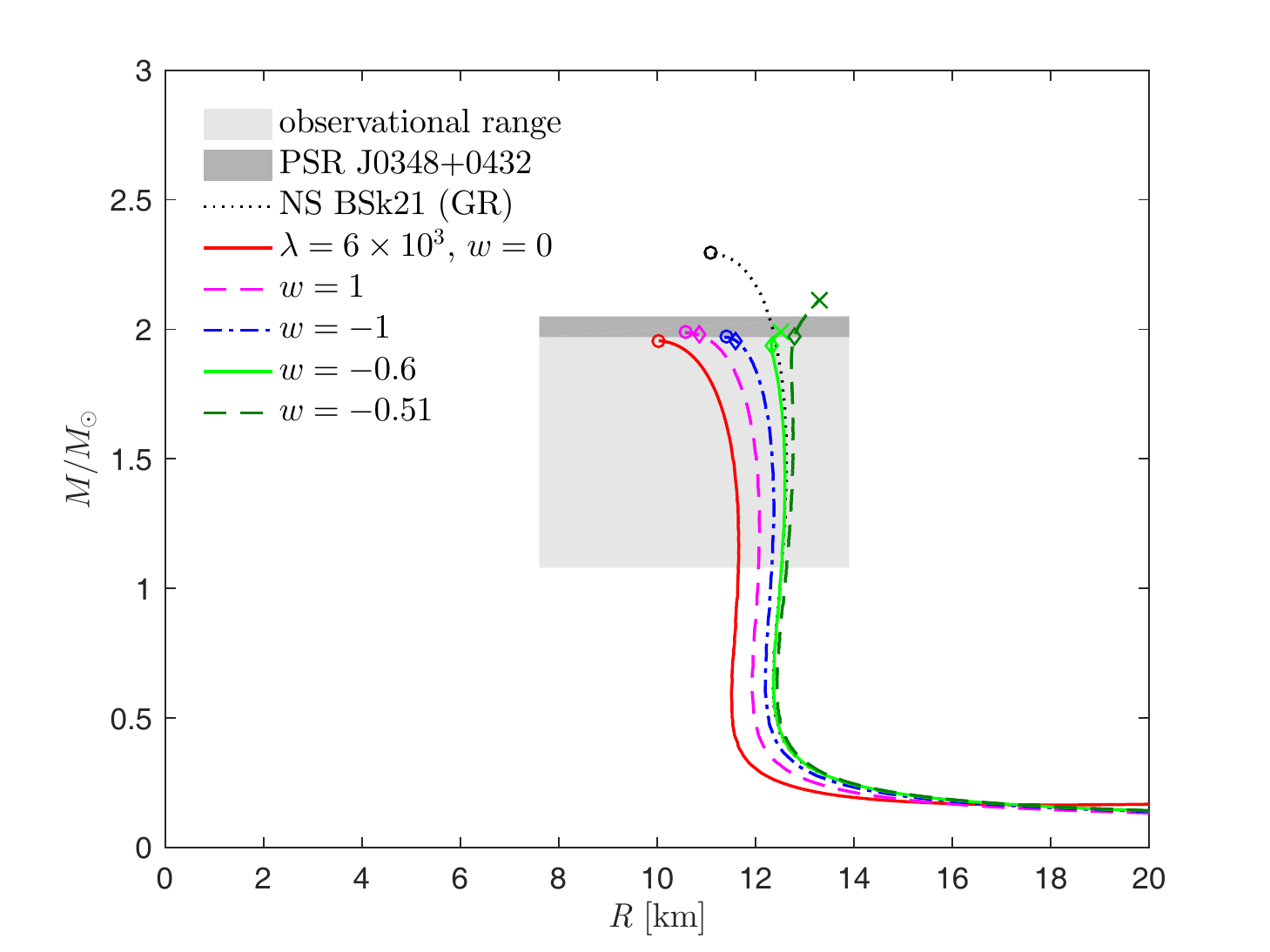}\\
		\hspace{-1em}\includegraphics[width=.5\textwidth]{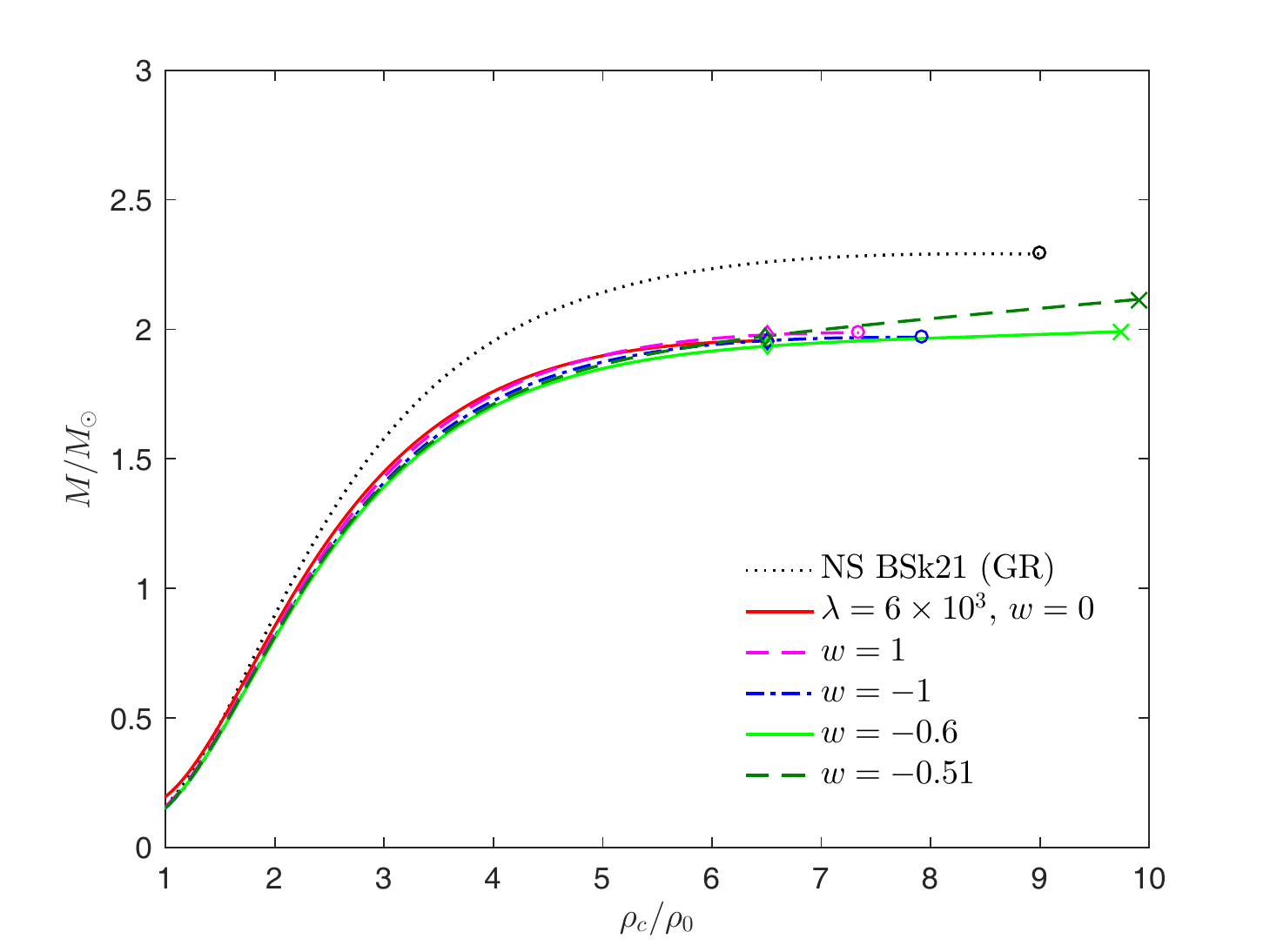}
	\end{tabular}
	\caption{\label{fig4} Top plot: Mass-radius relation for the  NS BSk21 EoS. The curves are given for different values of the dark EoS parameter $w$ and $\lambda=6\times 10^3$~MeV/fm$^3$. Bottom plot: The mass of the star versus the central energy density $\rho_c$ (in units of $\rho_0$).}
\end{figure}

We proceed to solve the system of brane TOV equations for different initial values of the central pressure $p_c$ in order to determine the mass-radius relation for such neutron stars. We consider first the case of a dark EoS with $w=0$. The results are presented in Figs.~\ref{fig3} and \ref{fig4}, where the stable mass-radius configurations, i.e. those that verify the stability criteria and obey the causality limit $v_{s,\mathrm{eff}} \leq 1$, are given. The shaded grey areas correspond to the observational constraints of Eqs.~\eqref{obsradii} and \eqref{obsmasses}, where the dark grey band is the mass range of the pulsar PSR J0348+0432~\cite{Antoniadis:2013pzd}. In the plots, the small circles denote the maximum mass $M_{max}$ (in units of the solar mass $M_\odot$) obtained by imposing the stability criterion alone, while the diamond markers indicate the maximum star mass at which the EoS subluminality bound $v_{s,\mathrm{eff}} \leq 1$ is violated. As $\lambda$ increases, this bound approaches the usual GR causality constraint $v_s \leq 1$, so that for larger values of $\lambda$ the stability condition is more restrictive than the causality bound. 

As can be seen from the top panel of Fig.~\ref{fig3}, the maximum mass and the corresponding star radius decrease as $\lambda$ decreases, so that requiring agreement with observational constraints leads to a lower bound on the brane tension, $\lambda \gtrsim 8\times 10^2$~MeV/fm$^3$. The star radii lie in the range 8--13~km. Furthermore, from the bottom panel of Fig.~\ref{fig4}, we conclude that $\rho_c\lesssim 9 \rho_0$. We also notice that the mass-radius curves bend clockwise for some negative values of $w$ (see e.g. the curves $w=-0.6$ and $w=-0.51$ in the top panel of Fig.~\ref{fig4}). In these cases, the maximum star mass is restricted by the speed-of-sound condition $v_{s,\mathrm{eff}} \leq 1$, since the stability condition is violated at much higher values of $M$ and $R$. For comparison, we have also indicated with crosses ($\times$) the mass-radius configuration at which the GR causality condition $v_s\leq 1$ is violated in such cases.

\begin{figure}[t]
	\centering
	\begin{tabular}{l}
		\hspace{-1em}\includegraphics[width=.5\textwidth]{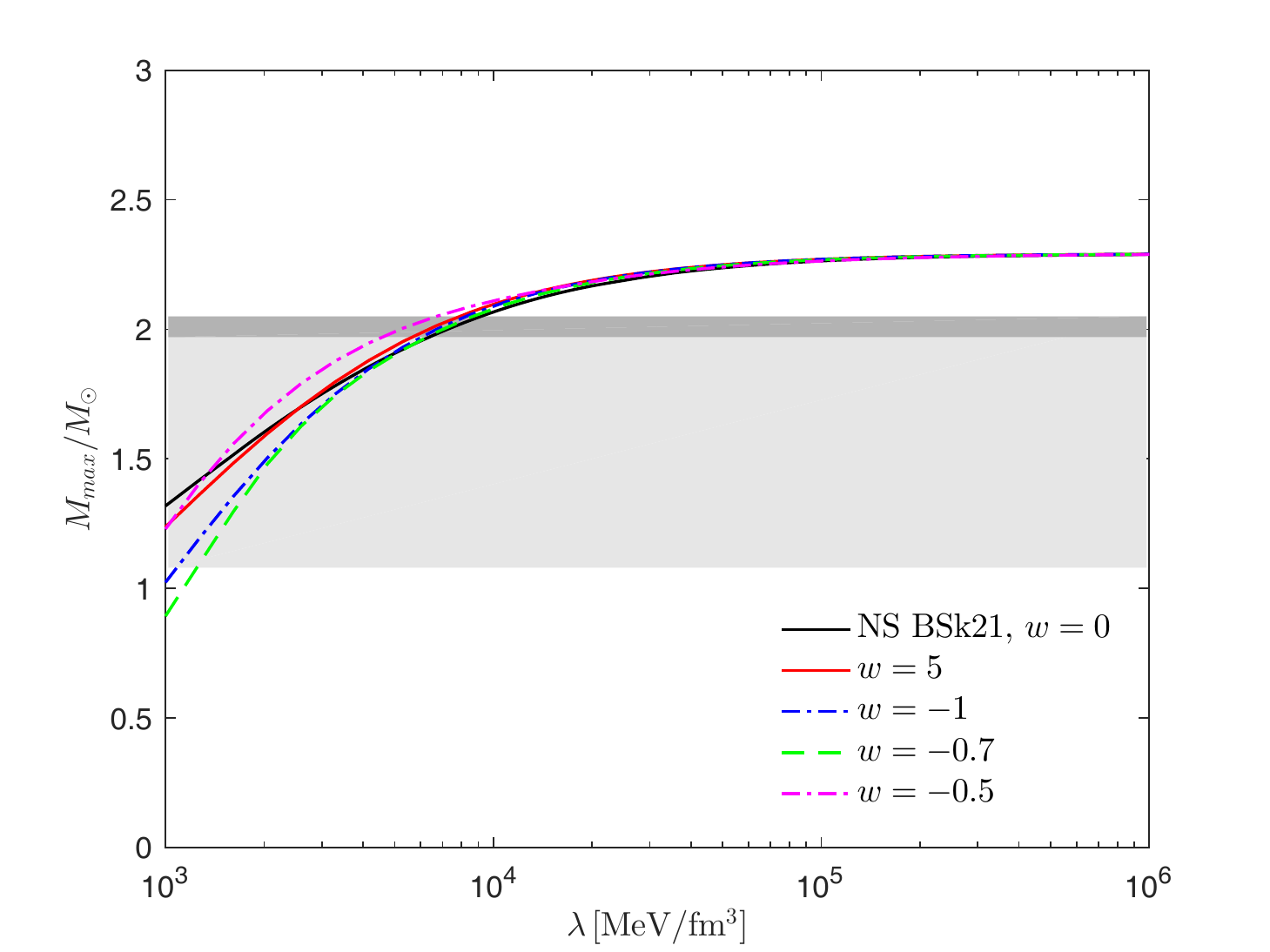}\\
		\hspace{-1em}\includegraphics[width=.5\textwidth]{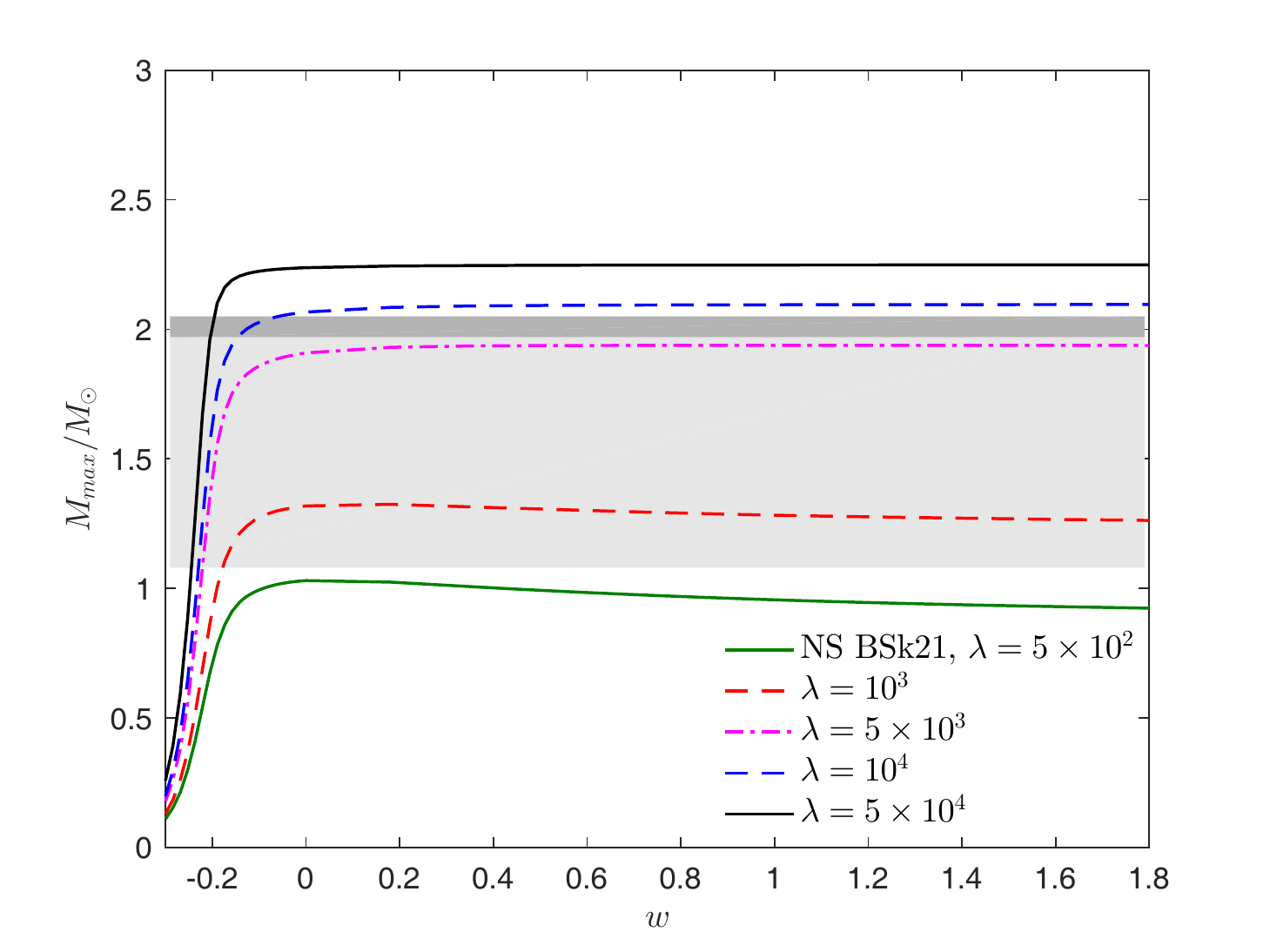}
	\end{tabular}
	\caption{\label{fig5} Top plot: Maximum neutron star mass as a function of the brane tension for different values of the dark EoS parameter $w$. Bottom plot: Maximum star mass as a function of the dark EoS parameter $w$ for different values of the brane tension.}
\end{figure}

In Fig.~\ref{fig5}, the maximum star mass is given as a function of the brane tension (top panel) and the dark EoS parameter $w$ (bottom panel). From the figure we conclude that the maximum star mass predicted for this type of EoS is compatible with observations provided that $\lambda\gtrsim 6\times10^2$~MeV/fm$^3$. For $w\gtrsim -0.1$, the value of $M_{max}$ remains practically constant with the variation of $w$, depending only on the value of $\lambda$. A similar behaviour is observed for $w \lesssim -0.5$. However, for $-0.3 < w <-0.1$, the maximum mass is quite sensitive to $w$, as it becomes evident in the bottom panel of Fig.~\ref{fig5}.

\subsection{Quark stars}

It has been conjectured that strange quark matter could be the true ground state of strongly interacting matter at zero pressure and temperature~\cite{Witten:1984rs}. If this possibility actually exists, it would open the window for the existence of compact stellar objects totally composed of SQM~\cite{Alcock:1986hz}. To describe the EoS inside such stars, we consider the simple phenomenological parametrisation given in Ref.~\cite{Alford:2004pf}. It consists of a power series expansion in the quark chemical potential $\mu$,
\begin{align}\label{QMeos}
\begin{split}
\rho &= \frac{9}{4\pi^2}a_4\mu^4-\frac{3}{4\pi^2}a_2\mu^2+B, \\
p &= \frac{3}{4\pi^2}a_4\mu^4-\frac{3}{4\pi^2}a_2\mu^2-B.
\end{split}
\end{align}

Besides the bag parameter $B$, this parametrisation contains two additional parameters, $a_4$ and $a_2$, which are independent of $\mu$ and are related to the QCD and strange quark mass (and color superconductivity) corrections, respectively. Note that for three-flavour quark matter consisting of free massless quarks one has $a_4=1$ and $a_2=0$. In this case, Eqs.~\eqref{QMeos} coincide with the well-known MIT bag model EoS, i.e. $\rho(p)=3p + 4B$~\cite{Chodos:1974je}. In our numerical calculations, however, we consider the more realistic values $a_4=0.7$ and $a_2=(180\,{\rm MeV})^2$~\cite{Alford:2004pf}. Furthermore, we take $B=60$~MeV/fm$^3$.\footnote{Assuming massless quarks and neglecting the strong coupling constant, the hypothesis that three-flavour quark matter has an energy per baryon lower than that of ordinary nuclei holds for $59~\text{MeV/fm}^3 \lesssim B \lesssim 92~\text{MeV/fm}^3$.}

\begin{figure}[t]
	\centering
	\begin{tabular}{l}
		\hspace{-1em}\includegraphics[width=.5\textwidth]{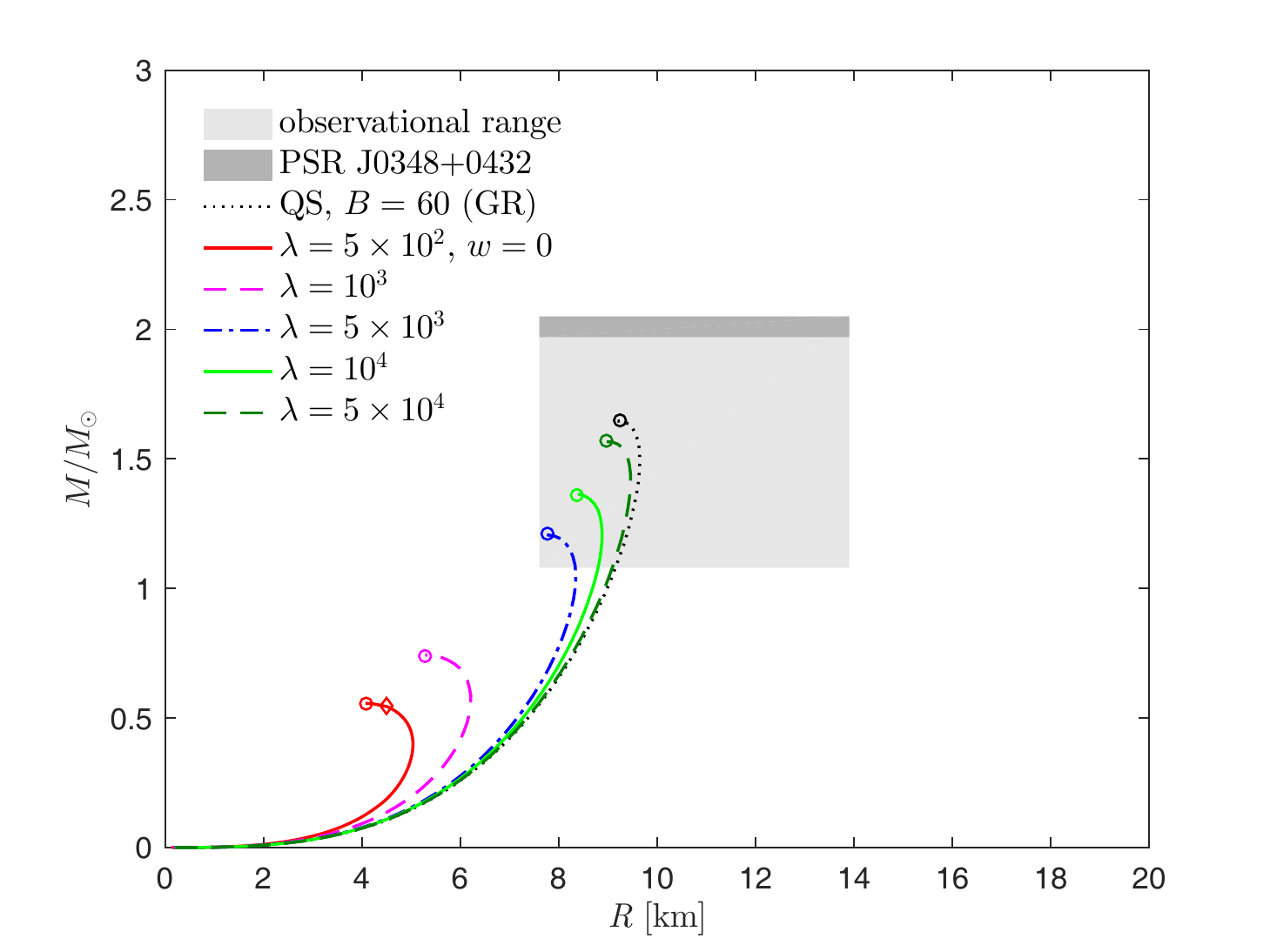}\\
		\hspace{-1em}
		\includegraphics[width=.5\textwidth]{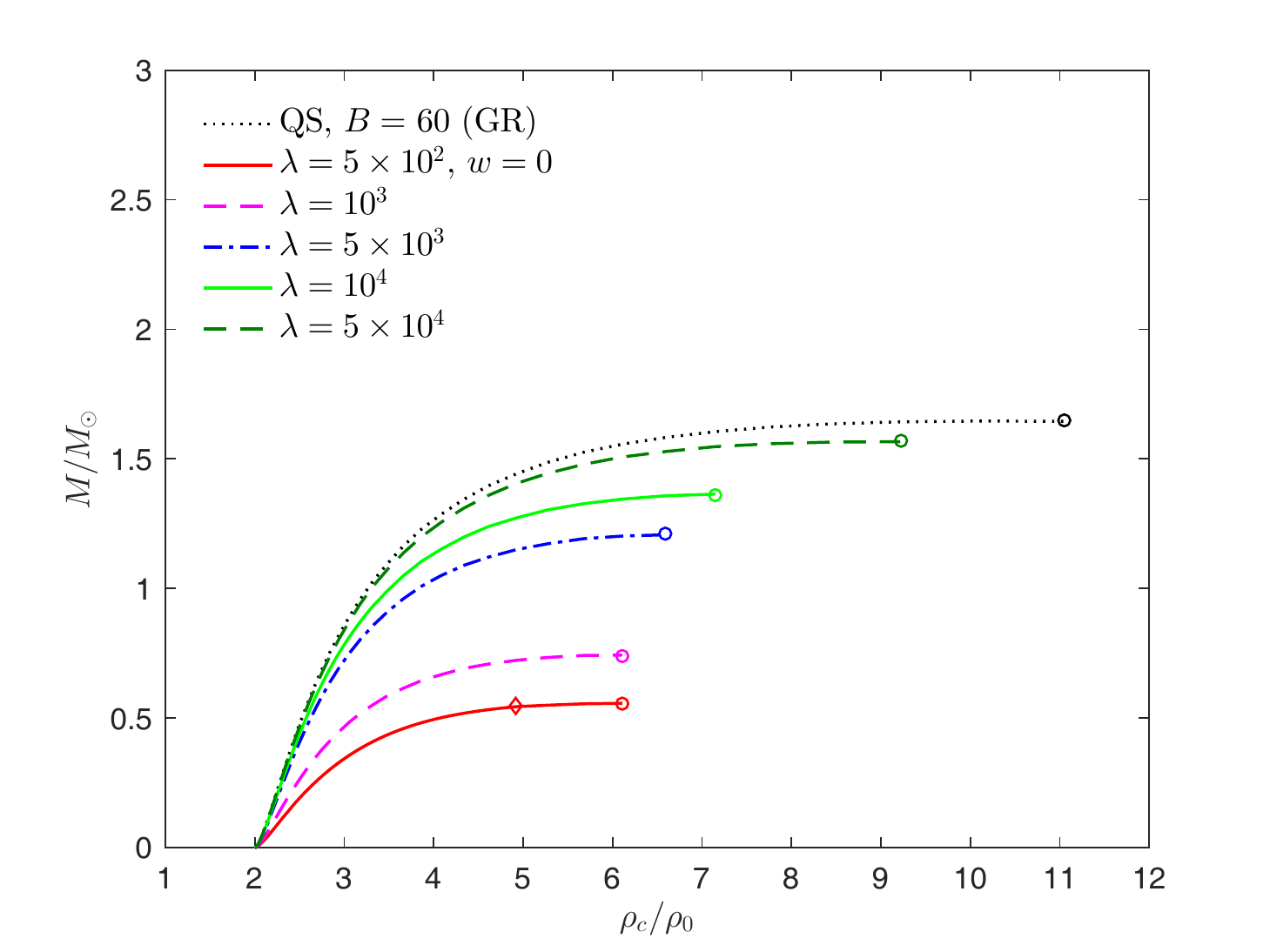}
	\end{tabular}
	\caption{\label{fig6} Top plot: Mass-radius relation for a quark star, using the phenomenological EoS in Eq.~\eqref{QMeos}. The curves are given for different values of the brane tension $\lambda$ (in MeV/fm$^3$), assuming a vanishing dark pressure in the star interior ($w=0$). Bottom plot: The mass of the quark star versus the central energy density $\rho_c$ (in units of $\rho_0$).}
\end{figure}

\begin{figure}[h]
	\centering
	\begin{tabular}{l}
		\hspace{-1em}\includegraphics[width=.5\textwidth]{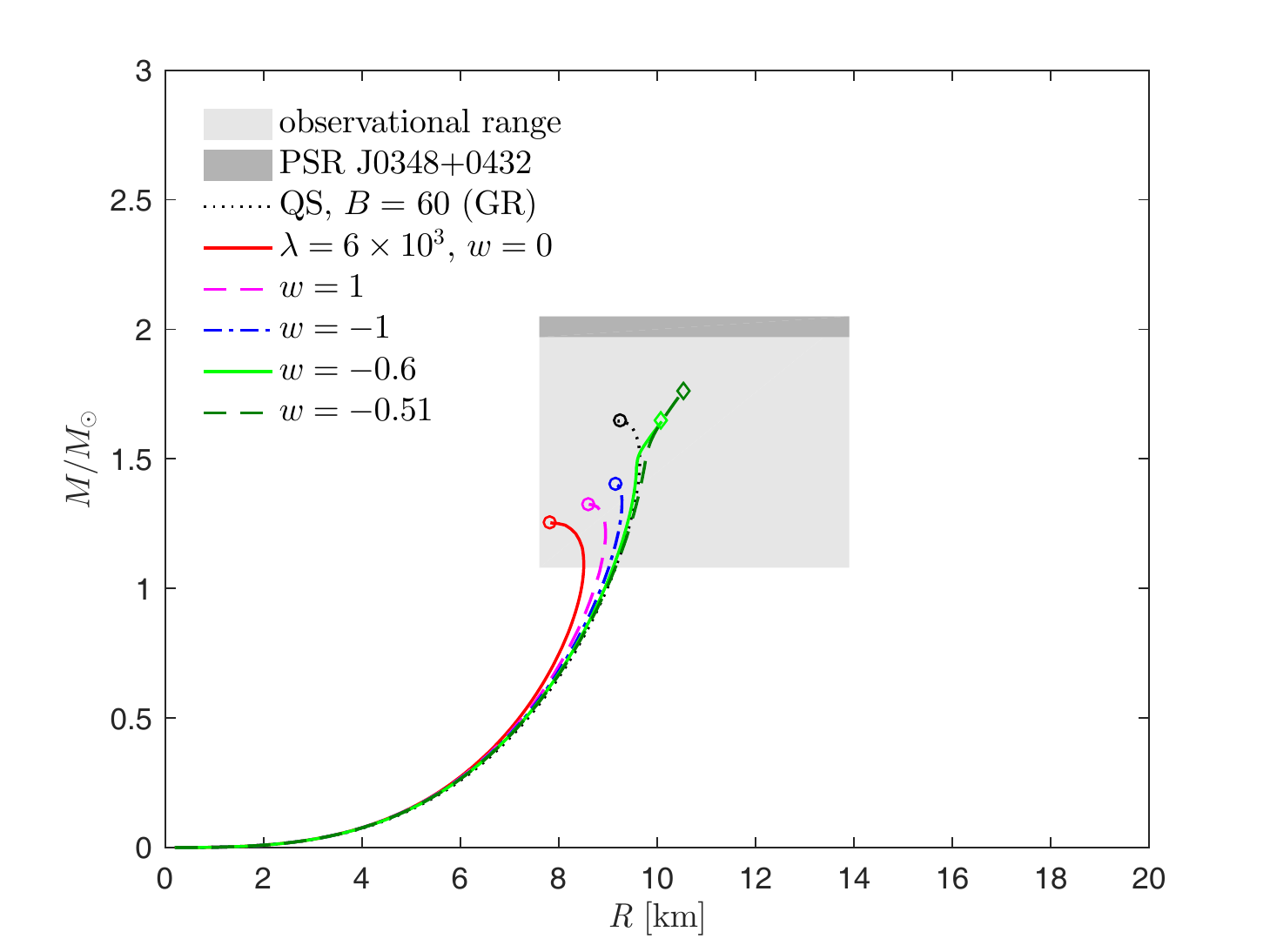}\\
		\hspace{-1em}
		\includegraphics[width=.5\textwidth]{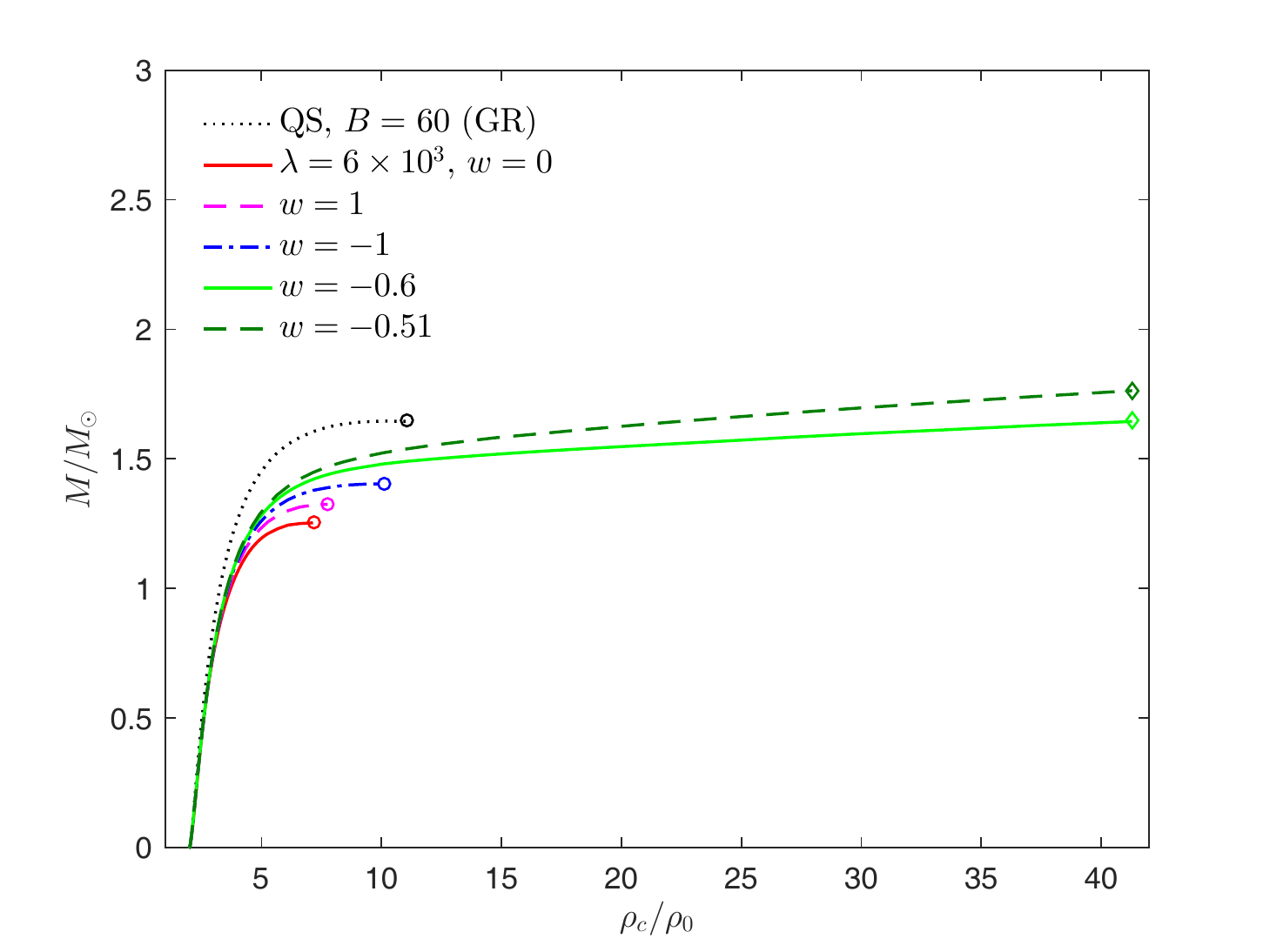}
	\end{tabular}
	\caption{\label{fig7} Top plot: Mass-radius relation for a quark star, using the phenomenological EoS in Eq.~\eqref{QMeos}. The curves are given for different values of the dark EoS parameter $w$ and $\lambda=6\times10^3$~MeV/fm$^3$. Bottom plot: The mass of the quark star versus the central energy density $\rho_c$ (in units of $\rho_0$).}
\end{figure}

We follow the same procedure as before, i.e. we integrate the brane-modified TOV equations for different initial values of the central pressure $p_c$ and determine the mass-radius relation for the corresponding quark star configuration. Our results are presented in Figs.~\ref{fig6} and \ref{fig7}. As shown in the top panel of Fig.~\ref{fig6}, the maximum mass and the corresponding star radius decrease as $\lambda$ decreases and agreement with observational constraints imposes the lower bound $\lambda \gtrsim 4\times 10^3$~MeV/fm$^3$, for $w=0$.  The allowed star radii are in the range 8--10~km. Moreover, from the bottom panel of the figure, we conclude that the central energy density for the maximum mass configurations is bounded by $\rho_c\lesssim 11\rho_0$.

In Fig.~\ref{fig7}, we present the mass-radius relation for the case of $\lambda=6\times10^3$~MeV/fm$^3$ and different values of $w$. We notice that for certain negative values of $w$ the mass-radius curves bend clockwise, reaching the maximum mass at relatively high central densities, $\rho_c \sim 40\rho_0$, bounded by the requirement of subluminality of the EoS. Although such density values are higher than the maximum central density of typical bare strange stars, they are  below the critical density required for the formation of a stable charm-quark star. We recall that a $c$-quark can be created via the weak reaction $u+d \rightarrow c+d$. Since the charm quark mass is $m_c \simeq 1.275$~GeV~\cite{Agashe:2014kda}, the production of a quark $c$ requires $\rho \geq \rho_{crit,c}=9m_c^4/(4\pi^2) \simeq 1.4\times10^{17}\, \text{g/cm}^3 \simeq 5.2\times 10^2 \rho_0$. Based on the stability analysis, it has been found in the context of GR that charm-quark stars are unstable against radial oscillations~\cite{Kettner:1994zs}. In the mass-radius plane, such an instability is manifested in the inwardly spiralling behaviour of the curves, and it is also confirmed through the calculation of the star oscillation frequencies~\cite{Kettner:1994zs}.

\begin{figure}[t]
	\centering
	\begin{tabular}{l}
		\hspace{-1em}\includegraphics[width=.5\textwidth]{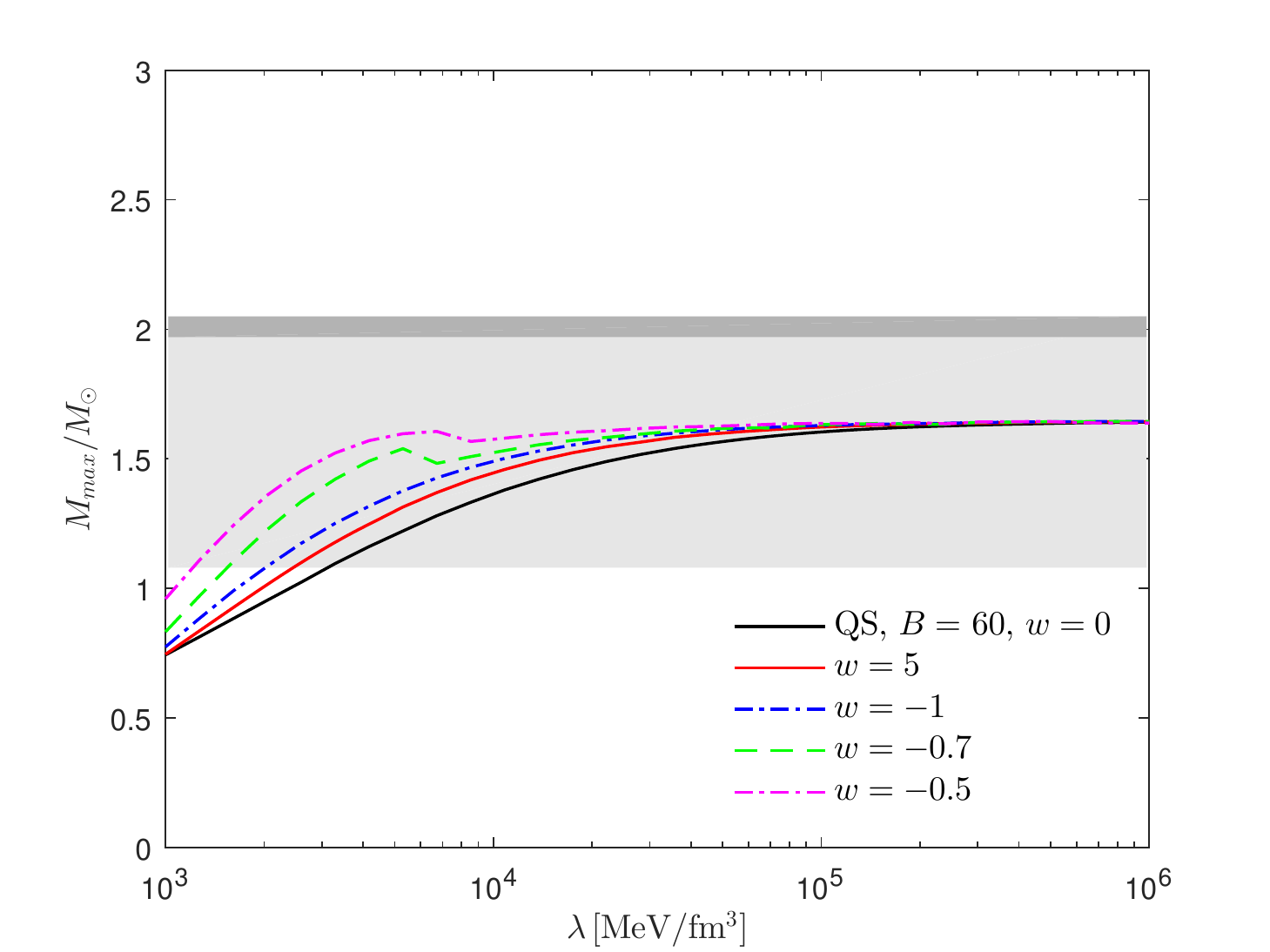}\\
		\hspace{-1em}\includegraphics[width=.5\textwidth]{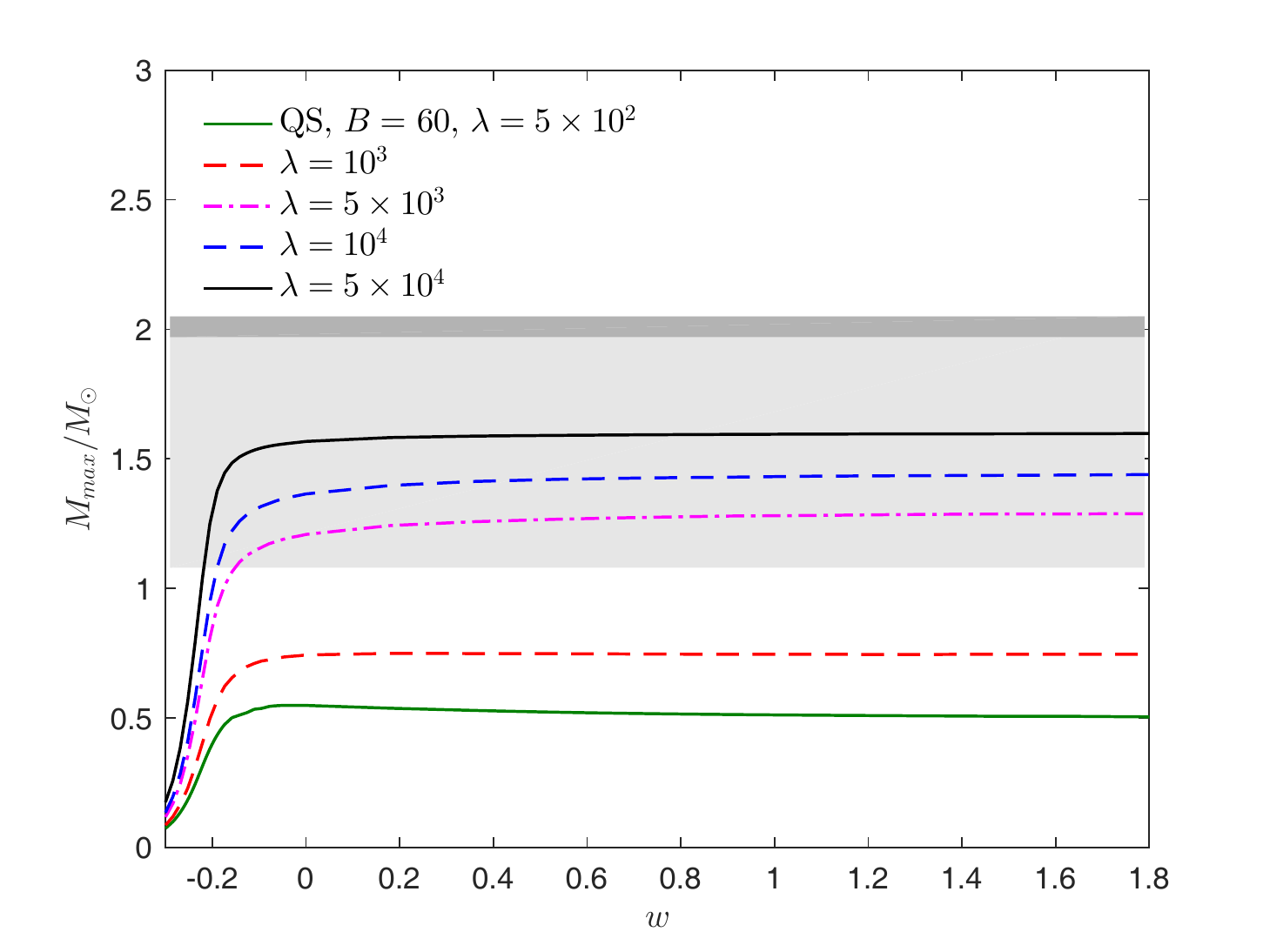}
	\end{tabular}
	\caption{\label{fig8} Top plot: Maximum quark star mass as a function of the brane tension for different values of the dark EoS parameter $w$. Bottom plot: Maximum star mass as a function of the dark EoS parameter $w$ for different values of the brane tension.}
\end{figure}

Finally, in Fig.~\ref{fig8}, the maximum star mass is presented as a function of the brane tension (top panel) and the dark EoS parameter $w$ (bottom panel). We see that the maximum star mass predicted for the quark model is compatible with observations provided that $\lambda\gtrsim 10^3$~MeV/fm$^3$. As in the case of pure neutron stars, for $w \gtrsim -0.1$, the value of $M_{max}$ remains essentially constant with the variation of $w$, and depends only on the value of $\lambda$. The same conclusion holds for $w \lesssim -0.5$. In the range $-0.3 < w <-0.1$, however, the maximum mass is quite sensitive to the value of $w$, as seen in the bottom panel of Fig.~\ref{fig8}.

\subsection{Hybrid stars}

The maximum values of the central energy densities $\rho_c$, obtained from EoS of purely hadronic matter, are typically in the range $5\rho_0$--$10\rho_0$. At such densities, one expects quark degrees of freedom to play a relevant role inside compact stars. It is then natural to consider the possibility of the formation of hybrid compact objects, i.e. stars that contain a core of quark matter with a crustal of nuclear matter. We describe a hybrid star using the following combined EoS. For densities $\rho \lesssim 4\rho_0$, the EoS is modelled by Eq.~\eqref{pnseos}, while for the core of the hybrid star, corresponding to densities $\rho > 4\rho_0$, we use the simple phenomenological parametrisation given in Eq.~\eqref{QMeos}. 

\begin{figure}[t]
	\centering
	\begin{tabular}{l}
		\hspace{-1em}\includegraphics[width=.5\textwidth]{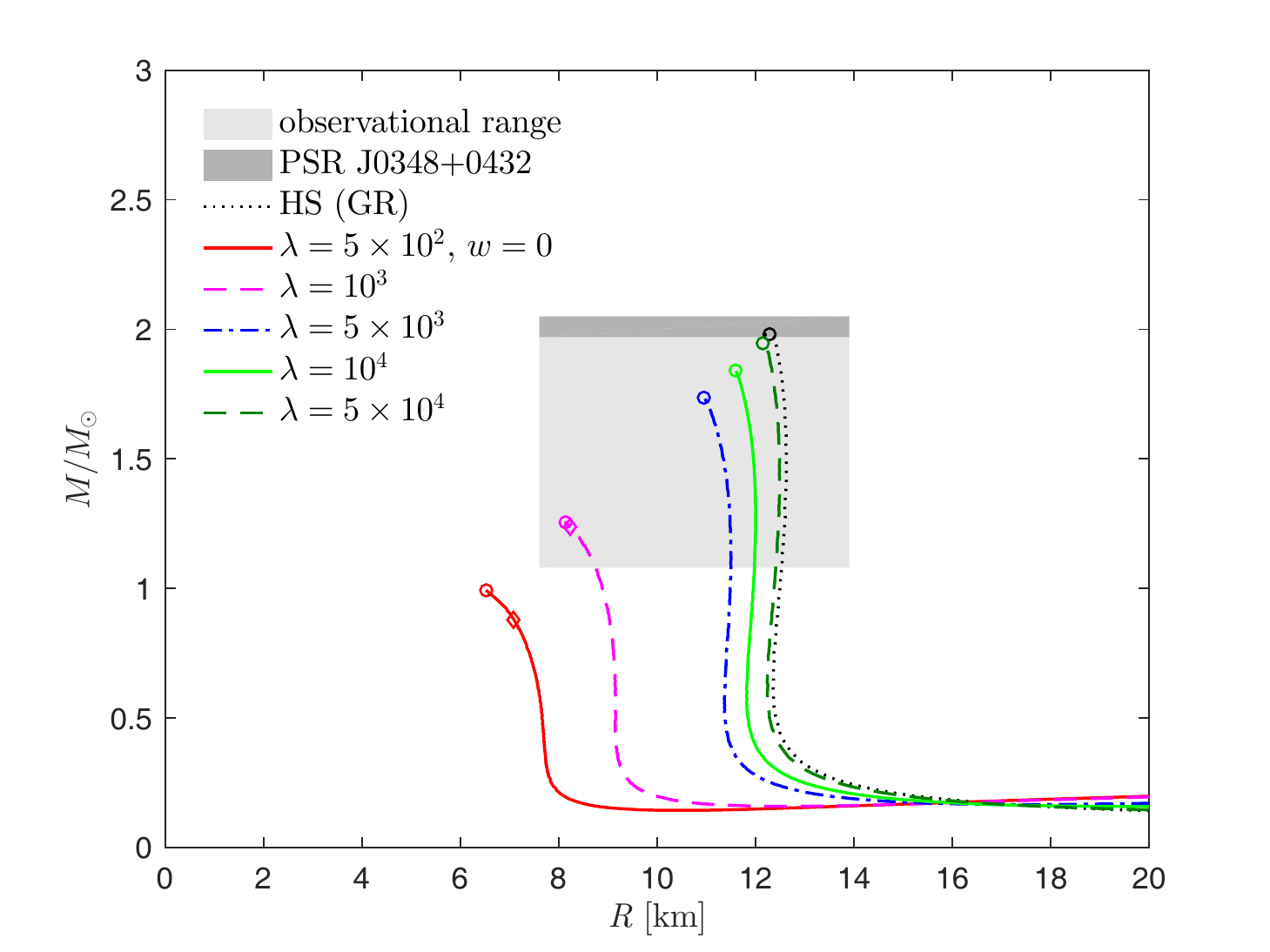}\\
		\hspace{-1em}
		\includegraphics[width=.5\textwidth]{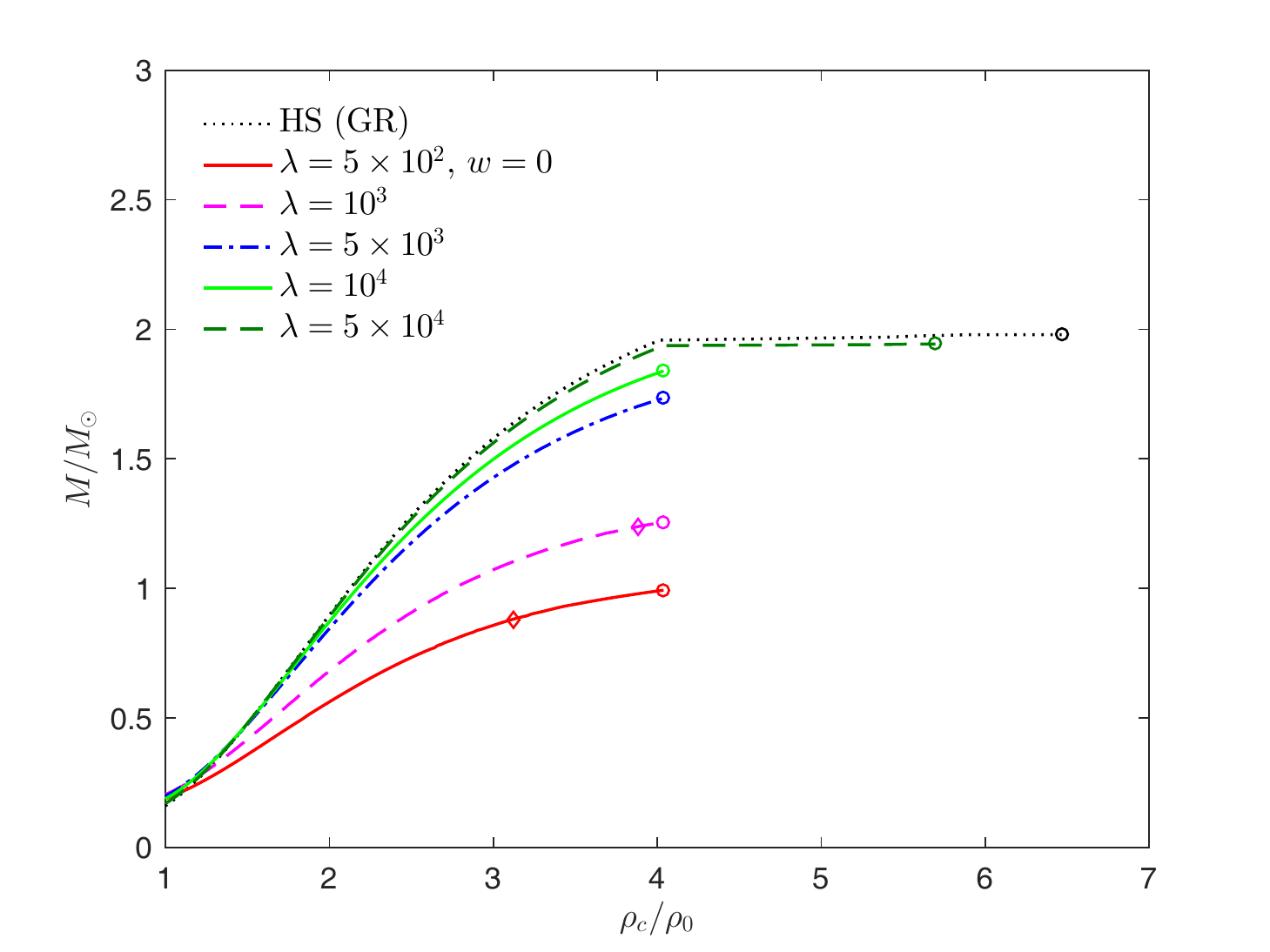}
	\end{tabular}	
	\caption{\label{fig9} Top plot: Mass-radius relation for the hybrid EoS. The curves are given for different values of the brane tension $\lambda$ (in MeV/fm$^3$), assuming a vanishing dark pressure in the star interior ($w=0$) Bottom plot: The mass of the star versus the central energy density $\rho_c$ (in units of the nuclear saturation density $\rho_0$).}
\end{figure}

\begin{figure}[h]
	\centering
	\begin{tabular}{l}
		\hspace{-1em}\includegraphics[width=.5\textwidth]{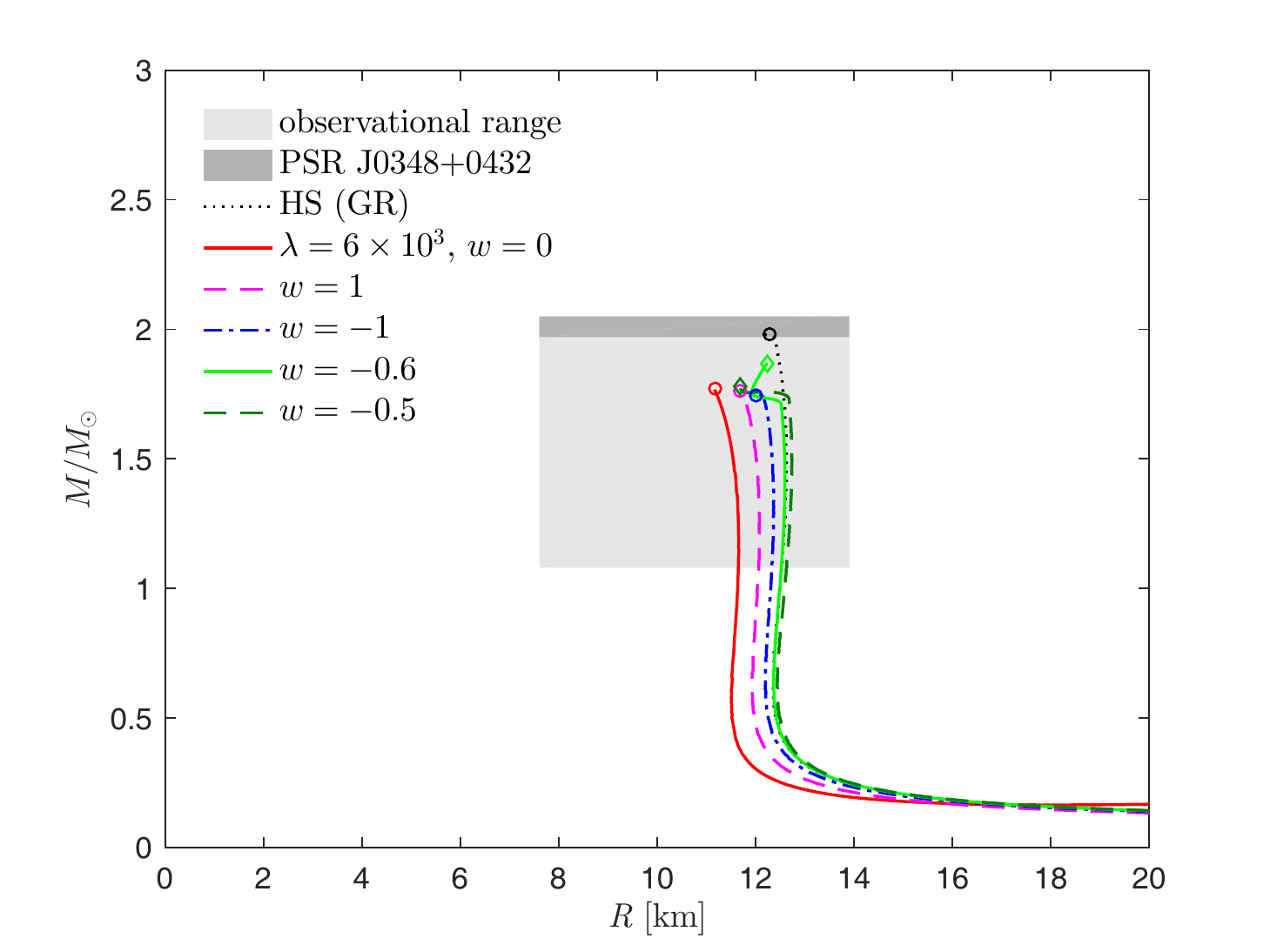}\\
		\hspace{-1em}
		\includegraphics[width=.5\textwidth]{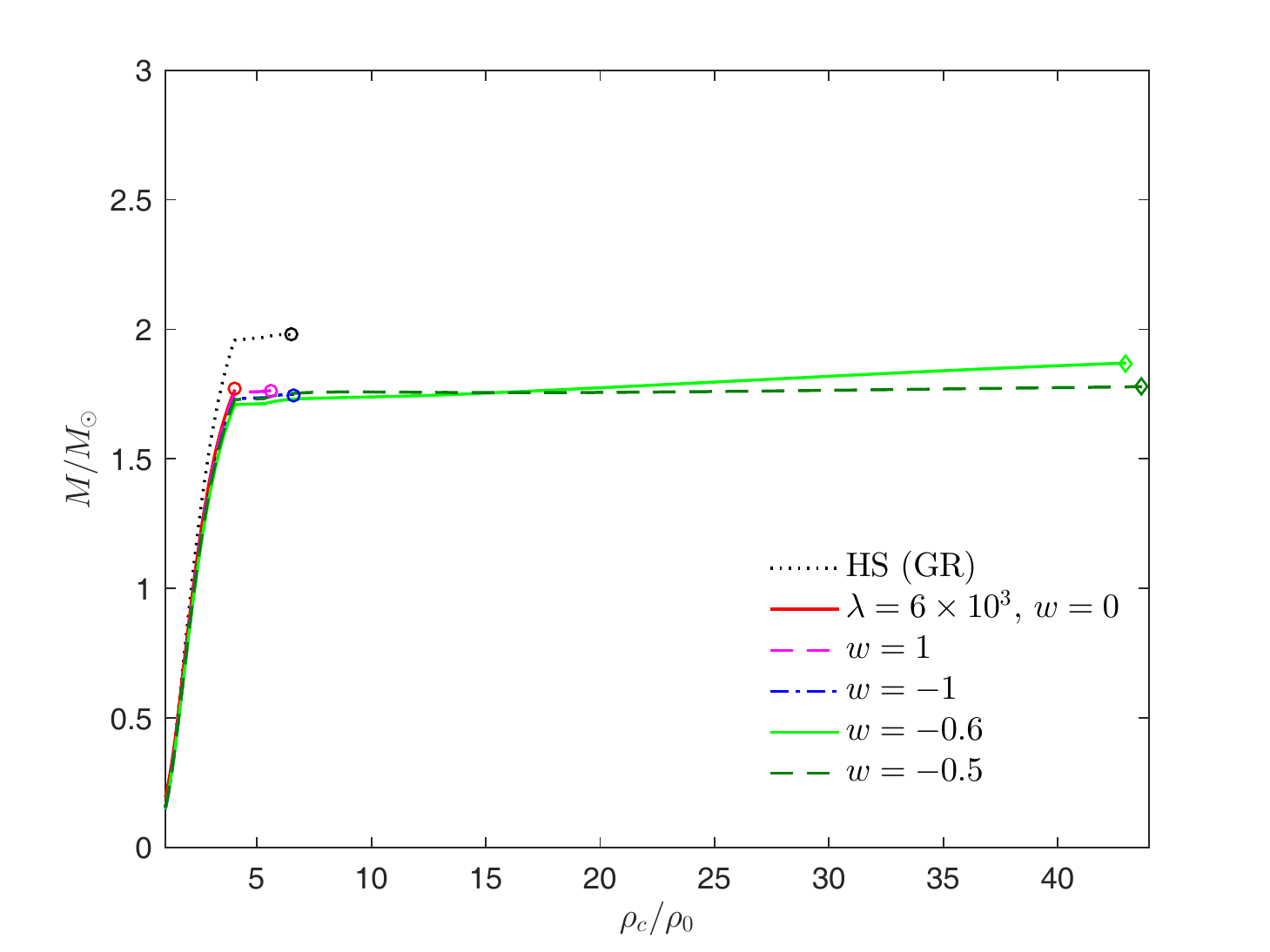}
	\end{tabular}
	\caption{\label{fig10} Top plot: Mass-radius relation for the hybrid EoS. The curves are given for different values of the dark EoS parameter $w$ and $\lambda=6\times 10^3$~MeV/fm$^3$. Bottom plot: The mass of the star versus the central energy density $\rho_c$ (in units of $\rho_0$).}
\end{figure}

We determine the mass-radius relation of the hybrid star by the integration of the brane-modified TOV equations for different initial values of the central pressure $p_c$.  Our results are presented in Figs.~\ref{fig9} and \ref{fig10}. From the top plot of Fig.~\ref{fig9}, we conclude that the maximum mass and the corresponding star radius decrease as $\lambda$ decreases. Furthermore, it is seen that for $\lambda \gtrsim8\times 10^2$~MeV/fm$^3$, and taking $w=0$, the masses and radii obtained are in agreement with the observational constraints. The maximum mass $M\sim 1.98 M_{\odot}$ is obtained for GR (i.e. in the limit $\lambda\to \infty$), and this value is consistent with the observational range of the pulsar PSR J0348+0432~\cite{Antoniadis:2013pzd}. This is to be compared with the maximum mass of $2.29 M_{\odot}$ attained with the NS BSk21 EoS (see Fig.~\ref{fig3}). The central energy densities are $\rho_c\lesssim 6.5\rho_0$ (bottom panel of Fig.~\ref{fig9}). The allowed star radii are in the range $8$--$13$~km, similar to those obtained for pure neutron stars.

In Fig.~\ref{fig10}, we present the mass-radius relation for the case of $\lambda=6\times10^3$~MeV/fm$^3$ and different values of the dark EoS parameter $w$. As in the case of quark stars, we notice that the clockwise bending of the mass-radius curves persists for certain negative values of $w$, reaching the maximum mass at relatively high central densities, $\rho_c \sim 45\rho_0$, as depicted in the bottom plot of Fig.~\ref{fig10}.

\begin{figure}[t]
	\centering
	\begin{tabular}{l}
		\hspace{-1em}\includegraphics[width=.5\textwidth]{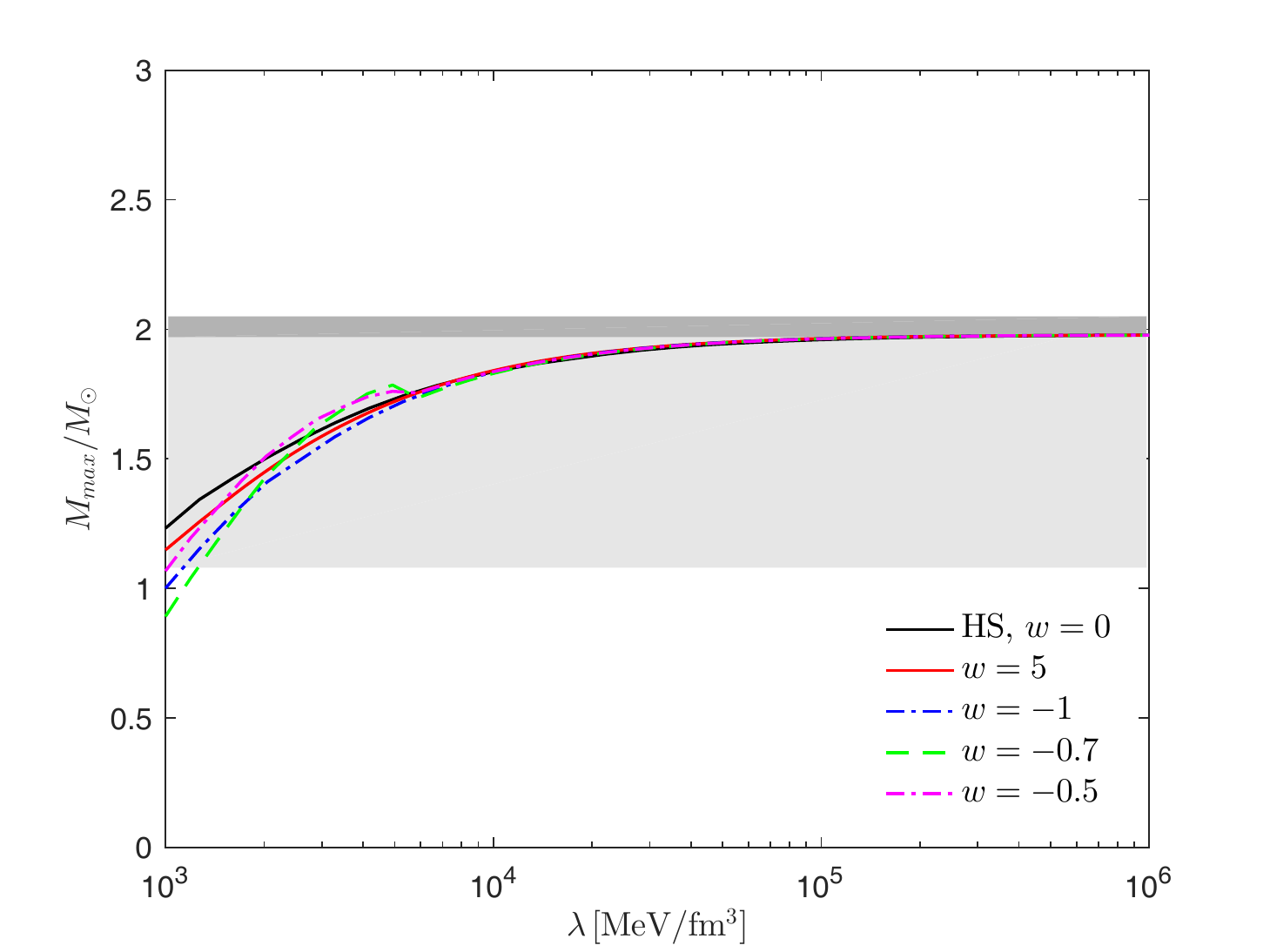}\\
		\hspace{-1em}
		\includegraphics[width=.5\textwidth]{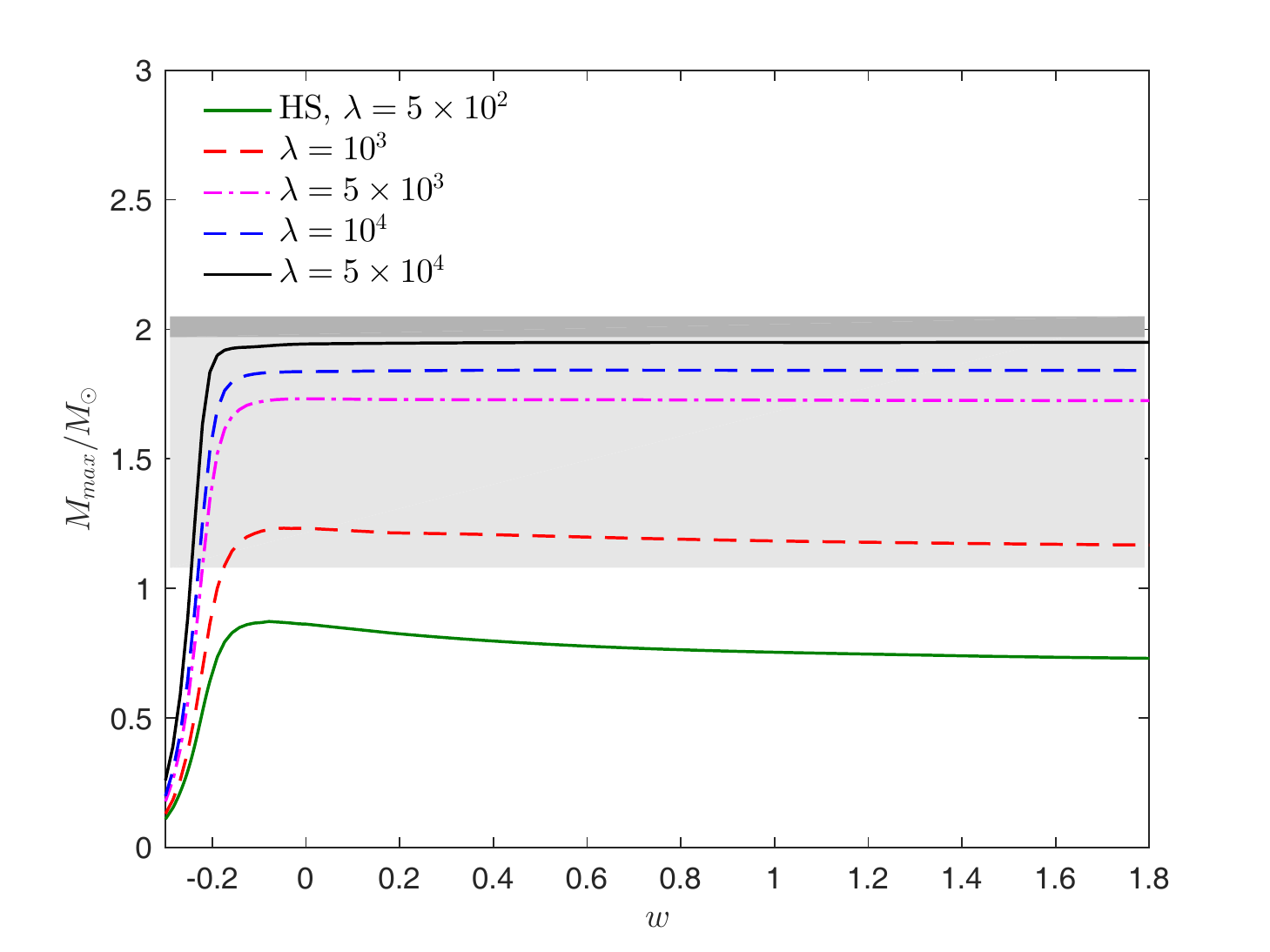}
	\end{tabular}
	\caption{\label{fig11} Top plot: Maximum hybrid star mass as a function of the brane tension for different values of the dark EoS parameter $w$. Bottom plot: Maximum star mass as a function of the dark EoS parameter $w$ for different values of the brane tension.}
\end{figure}

In Fig.~\ref{fig11}, the maximum star mass is plotted as a function of $\lambda$ (top panel) and $w$ (bottom panel). We see that the maximum star mass predicted for the hybrid star model is compatible with observations provided $\lambda \gtrsim 10^3$~MeV/fm$^3$. The behaviour of the maximum mass with $w$ is similar to the one obtained for pure neutron and quark stars. For $w\lesssim -0.5$ and $w \gtrsim -0.1$, the value of $M_{max}$ remains essentially constant with the variation of $w$, depending only on the value of $\lambda$. Yet, the maximum mass is very sensible to the variation of $w$ in the range $-0.3 < w <-0.1$ (as in the case of pure neutron stars). This is clearly seen in the bottom plot of Fig.~\ref{fig11}.

\section{Conclusions}
\label{sec:conclusions}

In this paper we have studied compact stars in the RS type-II braneworld. We have analysed the braneworld corrections to the different constraints for star masses and radii, considering compactness, causality and finite pressure at the core. We have solved the brane-modified TOV equations for different EoS that describe pure neutron stars, quarks stars and hybrid stars. We have then confront our results with recent astrophysical observations. 

To study the compactness limits and the effects of the brane corrections, we have solved the TOV equations assuming a star interior with $\mathcal{P}=\mathcal{U}=0$ and a uniform star density. We have shown that the brane corrections are significant for $\lambda \lesssim 10^4$~MeV/fm$^3$ and lead to a less compact star, when compared to the general relativity case. A $\lambda$-dependent limit was derived from the requirement of finiteness of pressure in the star core. The well-known compactness limits of general relativity are recovered in the limit $\lambda \rightarrow \infty$.

Significant deviations from the causality limit of general relativity are obtained when $\lambda \lesssim 10^4$~MeV/fm$^3$ in the case of $\mathcal{P}=0$ and $\mathcal{U}\neq 0$, i.e. for a dark EoS with $w=0$. In the latter case, the minimum radius is well approximated by a straight line whose slope varies with the brane tension. The constraints were obtained considering two different conditions for the speed of sound in dense matter, namely, $v_s\leq 1$ and $v_s\leq 1/\sqrt{3}$. We have also derived an effective speed of sound that includes corrections due to local and non-local braneworld effects. The causality condition $v_{s,\mathrm{eff}} \leq 1$ was then imposed on the different EoS in order to find stable stellar configurations.

The mass-radius relations were computed by solving the TOV equations for three different matter EoS, imposing the stability and causality criteria. The study was first done for $w=0$ and different values of the brane tension. In all the three EoS cases, the maximum mass and the corresponding star radius decrease as $\lambda$ decreases. Furthermore, the central energy density $\rho_c$ required to achieve the maximum mass configuration is always less than that of GR: $\rho_c \lesssim 9\rho_0$ (NS), $\rho_c \lesssim 11\rho_0$ (QS) and $\rho_c \lesssim 6.5\rho_0$ (HS). The star radii lie in the ranges $8$--$13$~km for neutron and hybrid stars, and $8$--$10$~km for quark stars.

The mass-radius relation was studied as a function of $w$ for the three types of compact stars and a given a value of the brane tension. As it turns out, the mass-radius curves exhibit in general a standard behaviour, i.e. they bend counterclockwise and the maximum mass is determined by the stability criteria. However, we have found that, in all three types of compact stars, the curves can bend clockwise for certain negative values of $w$, reaching the maximum mass at relatively high central densities for quark and hybrid stars: $\rho_c \sim 40\rho_0$ (QS) and $\rho_c \sim 45\rho_0$ (HS). In such cases, the maximum allowed mass is determined by the sound speed condition $v_{s,\mathrm{eff}} \leq 1$. 

Finally, the maximum star masses as functions of $\lambda$ and $w$ were studied for the three families of stars. Requiring agreement with observational constraints, a lower bound on the brane tension, $\lambda \gtrsim 10^3$~MeV/fm$^3$, is obtained for all three types of stars. The dependence of the maximum mass on $w$ are similar for the three EoS, namely, the maximum mass remains practically constant for  $w\lesssim -0.5$ and $w \gtrsim -0.1$, with the star mass being controlled by the brane tension $\lambda$. A remarkable feature in all EoS cases is the fact that, in the narrow range $-0.3 < w <-0.1$, the maximum star mass strongly depends on the value of the $w$ parameter.

\begin{acknowledgements}
The work of R.G.F. was partially supported by \emph{Associa\c c\~ao do Instituto Superior T\'ecnico para a Investiga\c c\~ao e Desenvolvimento} (IST-ID) and \textit{Funda\c{c}\~{a}o para a Ci\^{e}ncia e a Tecnologia} (FCT) through the projects CERN/FP/123580/2011 and UID/FIS/ 00777/2013. The work of A.P.M. and D.M.P. has been partially supported by the ICTP Office of External Activities through the project NET-35. R.G.F. thanks \emph{Instituto de Cibern\'etica, Matem\'atica y F\'{\i}sica} (ICIMAF) in Havana, Cuba, for hospitality. A.P.M. thanks \textit{Consejo Nacional de Ciencia y Tecnología} (CONACYT) for support through the sabbatical grant 264150 at ICN-UNAM, Mexico, where this work was finished.
\end{acknowledgements}

\bibliographystyle{spphys}
\bibliography{bibliografia}

\end{document}